\documentclass[11pt, a4paper]{article}

\usepackage[T1]{fontenc}
\usepackage{amsmath}
\usepackage{amssymb}
\usepackage{amsthm}
\usepackage{epsfig}
\usepackage{graphicx}
\usepackage{booktabs}
\usepackage{tikz}
\usetikzlibrary{snakes,fit,shapes}
\usepackage{url}

\newtheorem{thm}{Theorem}

\newtheorem{prop}[thm]{Proposition}

\newcommand{\mean}{{\mathbf E}}

\newcommand{\var}{{\rm var}}

\newlength{\ldescriptionwidth}
\newcommand{\ldescriptionlabel}[1]{
  \settowidth{\ldescriptionwidth}{{#1}}
  \ifdim\ldescriptionwidth>\labelwidth
    {\parbox[b]{\labelwidth}
      {\makebox[0pt][l]{#1}\\[1pt]\makebox{}}}
  \else
    \makebox[\labelwidth][l]{{#1}}%
  \fi
  \hfil\relax}

\newcommand{\note}[1]
{$^{(!)}$\marginpar[{\hfill\tiny{\sf{#1}}}]{\tiny{\sf{(!) #1}}}}

\def\doublespace{\baselineskip=\normalbaselineskip \multiply\baselineskip by 7
 \divide\baselineskip by 5}

\begin{document}

\doublespace

\title{On-line Spot Volatility-Estimation and Decomposition
with Nonlinear Market Microstructure Noise Models
%
%Particle Filter-Based On-Line Estimation of Spot Volatility with Nonlinear
%Market Microstructure Noise Models
%
\footnotetext{Address for correspondence: Institute of Applied Mathematics, University
of Heidelberg, Im Neuenheimer Feld 294, D-69120 Heidelberg, Germany (e-mail: jc@neddermeyer.net).
The work was supported by the University of Heidelberg under Frontier D.801000/08.023.} }
\author{Rainer Dahlhaus \\ \normalsize{(University of Heidelberg)} \\\\
Jan C. Neddermeyer \\\ \normalsize{(DZ BANK AG)}}
\date{July 2012}

\begin{titlepage}
\maketitle \thispagestyle{empty}

\noindent \textbf{Summary.} A technique for on-line estimation of spot volatility for
high-frequency  data is developed. The algorithm works directly on the transaction data and updates the volatility estimate immediately after the occurrence of a new transaction. Furthermore, a
nonlinear market microstructure noise model is proposed that reproduces several stylized facts of
high-frequency data. A computationally efficient particle filter is used that allows for the
approximation of the unknown efficient prices and, in combination with a recursive EM algorithm,
for the estimation of the volatility curve. We neither assume that the transaction times are
equidistant nor do we use interpolated prices. We also make a distinction between volatility per
time unit and volatility per transaction and provide estimators for both. More precisely we use a model with random time change where spot volatility is decomposed into spot volatility per transaction times the trading intensity - thus highlighting the influence of trading intensity on volatility.


\end{titlepage}

\setcounter{page}{1}

\newpage

%{\footnotesize \tableofcontents}

\section{Introduction}\label{ch:msn:intro}

In the last couple of years the modeling of financial data observed at high-frequency became one of
the major research topics in the field of financial econometrics. It is of high practical relevance
because a rising number of market participants execute trades based on high-frequency strategies
and are exposed to high-frequency market risk. Examples of those trading strategies are statistical
arbitrage, the execution of large block trades, and market making. For most strategies the spot
volatility is important for trading signal generation and risk management. Often the immediate
detection of sudden volatility movements is particularly relevant for traders. Usually
high-frequency trading strategies are highly automated. In fact, they are often ``speed games'' and
only profitable if one reacts to market changes faster than other market participants. An example is the pricing of high frequency options which can be traded until a few seconds to maturity. In a
high-frequency setting the estimation of spot volatility is much more complicated due to the
presence of market microstructure noise. Overall, this causes the need for an on-line spot
volatility estimator which filters out market microstructure noise and adapts to volatility
movements quickly. In addition, it needs to be computationally efficient. In this paper we propose such an estimation method.

%For high-frequency data the on-line estimation of
%spot volatility is a challenging task because of the presence of market
%microstructure noise.

In the method described below, the efficient log-price process of a security is treated as a latent
state in a nonlinear state-space model. The relation between the efficient log-prices and the
transaction prices is described by a class of nonlinear market microstructure noise models
leading to a particular form of the observation equation in the state-space model. A
computationally efficient particle filter is developed which allows the estimation of the filtering
distributions of the efficient log-prices given the observed transaction prices. Based on the
filtering distributions the time-varying volatility is estimated by using a  sequential
Expectation-Maximization (EM) algorithm. The procedure works on-line and updates the volatility
estimate immediately when a new transaction comes in. The method is suitable for real-time
applications because of its computational efficiency. Contrary to several other papers we do not
assume that the transaction times are equidistant nor do we use interpolated prices.

Until recently, the main focus in the literature has been on the estimation of the integrated
volatility. This task has been studied extensively under various assumptions on the market
microstructure noise (Zhou 1996; Zhang et al. 2005; Andersen et al. 2006; Bandi and Russell 2006,
2008;  Hansen and Lunde 2006; Barndorff-Nielsen et al. 2008; Kalnina and Linton 2008; Christensen
et al. 2009; Jacod et al. 2009; Podolskij and Vetter 2009). Some authors suggested that estimates
of the spot volatility can be obtained through localized versions of estimators for the integrated
volatility (Harris 1990; Zeng 2003; Fan and Wang 2008; Bos et al. 2009;
Kristensen 2010) or by Fourier series methods (Munk and Schmidt-Hieber 2009). A specific noise-robust estimator is provided with a detailed analysis by Zu and Boswick (2010).
However, these
methods are essentially off-line procedures. Foster and Nelson (1996) derive the rate of convergence of rolling regression estimates. They include the case of one-sided kernel-estimates which can be transformed to recursive estimates.

In this article we use a diffusion model with random time change given by the number of transactions (see Section~\ref{ch:msn:clock}). Conditional on the observed trading times $t_1 < t_2 < \ldots < t_T$ the evolution of the unobserved efficient log-price process $X_{t_j}$ is given by a random walk in transaction time with possibly time-varying volatility
$\sigma_{t_j}$, that is
\begin{equation}\label{rounding:stateequation}
X_{t_j} = X_{t_{j-1}} + Z_{t_j}
\end{equation}
with $Z_{t_j} \!\sim\! \mathcal{N}(0,\sigma_{t_j}^2)$ (an alternative is a diffusion model in clock
time - see Section~\ref{ch:msn:clock}). Drift terms are ignored since their effect is of lower
order in high-frequency data. The observed transaction data $Y_{t_j}$ are then treated as noisy observations of the latent process $X_{t_j}$.

In Section~\ref{ch:msn:clock}, a transformation from transaction time volatility $\sigma_{t_j}^2$ to clock time volatility is given. The basis for this is that the underlying diffusion model with random time change leads to a decomposition of volatility in clock time into volatility in transaction time and trading intensity. This shows in particular the influence of the local trading intensity on volatility. In addition, we present a direct clock time estimator.

In our opinion the main advantage of the above model in comparison with a continuous time diffusion model is the aforementioned decomposition of the volatility discussed in Section~\ref{ch:msn:clock}. In addition, volatility in transaction time is more constant than volatility in clock time making the algorithm more stable (An{\'{e}} and Geman 2000; Plerou et al. 2001; Gabaix et al. 2003 -- see also Section~\ref{ch:msn:sim:realdata}).

The relation between the unobserved efficient (log-)prices and the observed
transaction prices is described through a nonlinear market microstructure noise model given by the generalized rounding scheme
\begin{equation}\label{rounding:obsequation}
Y_{t_j} = g_{t_j}\!\big(\exp(X_{t_j})\big) = g_{t_j;Y_{t_{1:j-1}}
}\!\big(\exp(X_{t_j})\big).
\end{equation}
Here the function $g_{t_j}$ may be random or deterministic, time-inhomogeneous, and depending on the past observations $Y_{t_{1:j-1}} := \{ Y_{t_1}, \ldots, Y_{t_{j-1}} \}$ and in addition on exogenous variables such as order book data or market maker quotes. Contrary and complementary to the additive model $\log Y_{t_j} = X_{t_j} + U_{t_j}$ used in the majority of
existing papers this model tries e.g. to describe in detail the rounding-mechanism due to order books and market maker quotes.

The particle filter applied in this paper allows for a fairly general class of nonlinear functions $g_{t_j}\!\big(\cdot)$ possibly depending on unknown parameters (cf.(\ref{alternat:msnmodel}) for an example different to the mainstream of this paper) - for computational simplicity and since we believe that this is a very good model we restrict ourselves to the setting (\ref{modelass1}) where the possible support of $\exp(x_{t_j})$ can be diagnosed from $y_{t_j}$, previous observations and (say) the order book. A simple deterministic example covered by this model is the rounding of $\exp(X_{t_j})$ to the nearest cent. A more complex stochastic example is the situation where the next trade is made both with probability $1/2$ on the closest bid- or ask-level of an order book. One might be tempted to write our model in the form
$Y_{t_j} = \tilde{g}_{t_j}\!\big(\exp(X_{t_j}), U_{t_j}\big)$ with a deterministic nonlinear $\tilde{g}_{t_j}$ and a random component $U_{t_j}$ - but in most situations the $U_{t_j}$ would depend in a very complicated way on the past of the $Y_{t_j}$ (e.g. in Example 4 from Section~\ref{ch:msn:msnmodel}).

The state equation (\ref{rounding:stateequation}) and the observation equation
(\ref{rounding:obsequation}) form a nonlinear state-space model (see also (\ref{modelass1}) and
(\ref{distributional_state-equation})). The (transaction time) spot volatility curve is considered as a parameter of
this state-space model. The estimation is done through a particle filter and a sequential
EM-type algorithm. Very roughly speaking the volatility estimator can be viewed as a localized
realized volatility estimator based upon the particles of the particle filter. In detail the
situation is however more complicated because we need a back and forth between particle filter and
volatility estimator to obtain a decent on-line estimator.

%We mention that our methods are not restricted to the above model but can also
%be applied with other microstructure noise models. Contrary
%to several other papers we do not assume that the transaction times are equidistant
%nor do we use interpolated prices.

The article is organized as follows. Section~\ref{ch:msn:msnmodel} describes the nonlinear market
microstructure noise model. In Section~\ref{ch:msn:ssmEMalg}, a particle filter and a sequential
EM-type algorithm are proposed for on-line estimation of spot volatility in transaction time.
In Section~\ref{ch:msn:clock} the decomposition of clock time volatility is given and the estimation of spot volatility in clock time is discussed - both in a diffusion model with random time change and in a standard diffusion model. Modifications (e.g. for data with diurnal patterns), adaptation issues, and the implementation of the algorithm are discussed in Section~\ref{ch:msn:impl}. Finally, simulation results
and an application to real data are presented in Section~\ref{ch:msn:sim} followed by some
conclusions in Section~\ref{ch:msn:concl}.

In some parts of the paper we could replace the notation $y_{t_j}$, $x_{t_j}$ by the simpler
notation $y_{j}$, $x_{j}$ etc. Since the time points $t_j$ are treated in Section~\ref{ch:msn:clock} as the realization of a point process we have decided to stick with this notation throughout.

\section{Nonlinear Market Microstructure Noise Models}
\label{ch:msn:msnmodel}

In most existing market microstructure models the efficient log-price is assumed to be corrupted by
additive stationary noise (cf. A\"{i}t-Sahalia et al. 2005; Zhang et al. 2005; Bandi and Russell 2006;
Hansen and Lunde 2006; Barndorff-Nielsen et al. 2008). The noise variables are typically
independent of the efficient log-price process. The setting allows for weak (even nonparametric) assumptions on the noise and fairly general theoretical results of the estimators. The major weakness of these models is that they cannot reproduce the discreteness of transaction prices. More adequate models which incorporate rounding noise have also been considered (Ball 1988; Delattre and Jacod 1997; Large 2007; Li and Mykland 2007; Robert and Rosenbaum 2008; Rosenbaum 2009). Popular models are based on additive noise followed by rounding
according to the smallest tick size as in (\ref{alternat:msnmodel}). At the end of Section~\ref{ch:msn:ssmEMalg:pf} we discuss how these models can be used in the framework of our paper. Hansen and Horel (2009) use a Markov chain model for the filtering of discretized prizes and consider realized volatility estimates based on the filtered series.

As already described in (\ref{rounding:obsequation}) the observed price in our model is obtained from the unknown efficient price by application of the generalized rounding function $g_{t_j}$ which may be deterministic or stochastic - examples are simple deterministic and stochastic rounding (Example 1 below), rounding to the closest liquid bid or ask levels of an order book (Example
2), rounding to the quotes created by a market maker (Example 3) or to
some levels estimated from previous observations $Y_{t_{1:j-1}}$ (Example 4) where $g_{t_j}$ depends in a complicated way on past observations. In general the rounding levels are
created from additional exogenous information or from previous observations
$y_{t_{1:j-1}}$ (e.g. the levels in an order book). In most cases we assume that (conditional on
$y_{t_{1:j-1}}$) these levels are known. Example 4 is an example where such exogenous information
is unavailable and the levels are estimated. Examples 1-4 do not contain unknown parameters (although this is not mandatory - cf. (\ref{alternat:msnmodel}) which depends on $\sigma_{U}^{2}$).

%From a theoretical (state-space) point of view it would be natural to either define directly the
%rounding function $g_{t_j;y_{t_{1:j-1}}}(\cdot)$  or, even better, to specify the distribution of
%the observation conditional on the state $p\big(y_{t_j}\big| y_{t_{1:j-1}}, \exp(x_{t_j})\big)$ as
%it is done in (\ref{modelass1}) (note that $Y_{t_j}$ is a discrete random variable and $A_{t_j}$
%depends on $y_{t_j}$). In combination with (\ref{distributional_state-equation}) this is the basis
%for the particle filter. This is the standard approach which explains how the observation is
%created from the state. For financial transaction data the situation is often vice versa: using the
%observation one draws conclusions on the underlying state (cf. Example 3). This is the reason why
%we base the following definition on the inverse image of the function
%$g_{t_j;y_{t_{1:j-1}}}(\cdot)$. (\ref{modelass1}) then follows immediately.

\bigskip

\noindent \textbf{Model assumption 1 (observation equation / microstructure noise model):}

\medskip

\noindent (i) \; The distribution of $Y_{t_j} = g_{t_j;Y_{t_{1:j-1}}
}\!\big(\exp(X_{t_j})\big)$ is discrete with support $\mathcal{Y}$.\\[6pt]
(ii) The conditional distribution of $Y_{t_j}$ given the state $X_{t_j}$ and previous
observations is\linebreak \hspace*{0.65 cm} of the form
\begin{equation} \label{modelass1}
p\big(y_{t_j}\big| y_{t_{1:j-1}}, \exp(x_{t_j})\big) \propto
\mathbf{1}_{A_{t_j}}\big(\exp(x_{t_j})\big) \, \mathbf{1}_{\mathcal{Y}} (y_{t_j})\qquad \mbox{a.s.}
\end{equation}
\hspace*{0.65 cm} where the set $A_{t_j}$ depends on $y_{t_j}$ and on the conditioning observations
$y_{t_1},\ldots,y_{t_{j-1}}$.\linebreak \hspace*{0.65 cm} ``a.s.''  means almost surely with
respect to the distribution of $X_{t_j}$.

\bigskip

\noindent It is important to note that the concrete specification of the set $A_{t_j}$ is an
important part of the microstructure noise model at hand - see Examples 1-4 below. Note that we need to know  $A_{t_j}$ only
for the observed $y_{t_j}$ and not for all possible realizations of $Y_{t_j}$. If the function
$g_{t_j}=g_{t_j;Y_{t_{1:j-1}}}$ is deterministic then $A_{t_j} := g_{t_j}^{-1}(y_{t_j}) = \{z :
g_{t_j}(z)= y_{t_j}\}$ is the inverse image of $y_{t_j}$ under $g_{t_j}$. The ``a.s.'' will be
omitted in the rest of the paper. In particular Proposition~\ref{msn:prop1} continues to hold if
(\ref{modelass1}) only holds almost surely.

\medskip

\begin{figure}[ht]
\centering
\includegraphics[width=410pt,keepaspectratio]{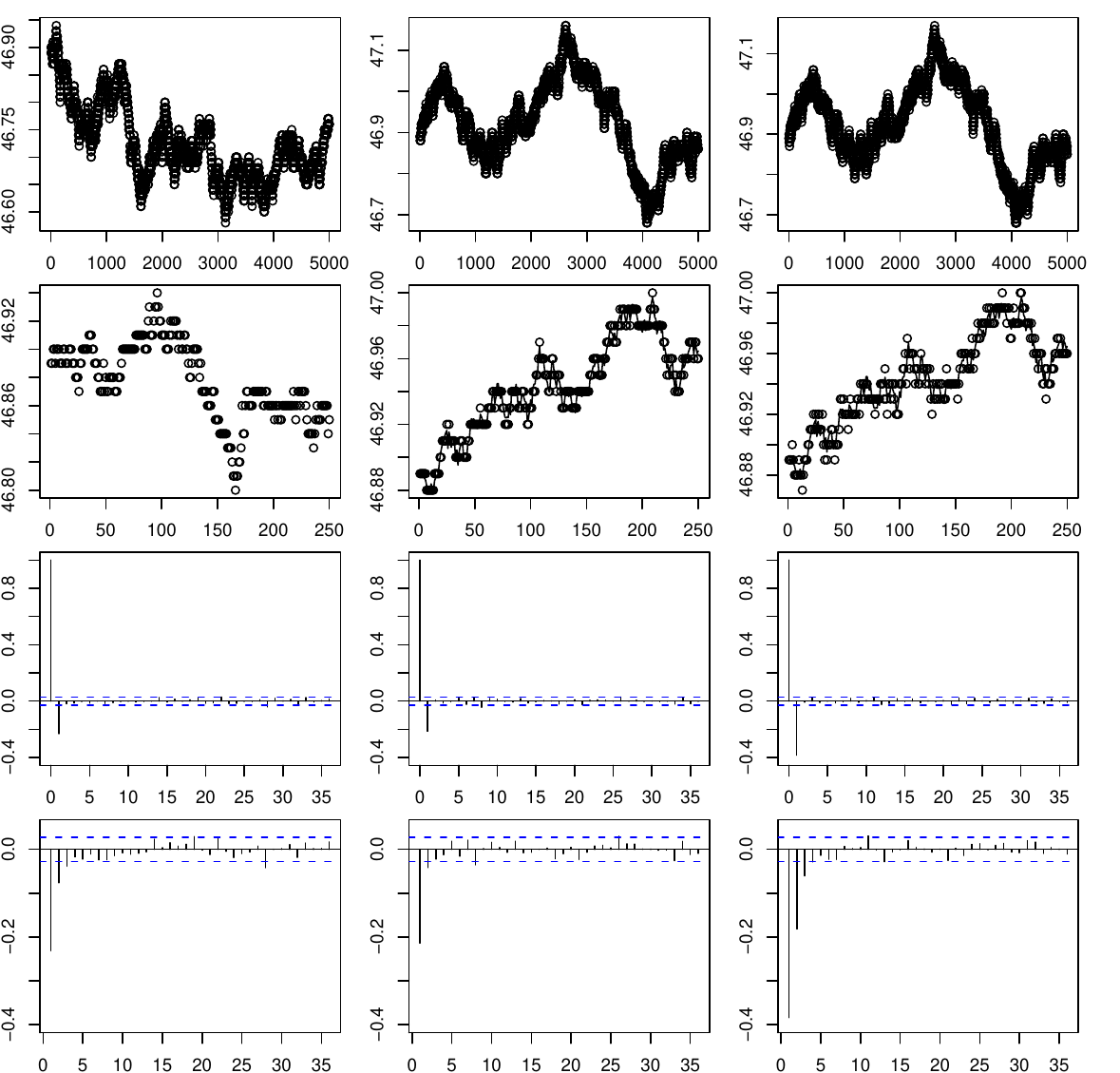}
\caption{\footnotesize
Comparison of real transaction data for Citigroup (left column) with simulated
data from the market microstructure noise model with deterministic rounding (middle column) and stochastic rounding (right
column). The plots show (from top to bottom): 5,000 transaction prices; the
first 250 transaction prices and the efficient price process of the simulated
data; the autocorrelations and partial autocorrelations of the returns of
the transaction prices.}
\label{fig:SimulatedvsRealData}
\end{figure}

\noindent \underline{Reasons for choosing the model:} \; We chose the above model for three reasons:\\[6pt]
(a) The particle filter takes a simple form: The optimal proposal becomes a truncated normal
distribution and the importance weights are also easy to calculate (see Proposition~1). This means
that the filter is more efficient than in the general case and less particles (and less
computation time) are
needed.\\[6pt]
(b) The model covers several important cases of microstructure noise (see the Examples below).\\[6pt]
(c) Already the simplest model of deterministic rounding in Example 1 (in combination with order book data or market maker quotes) describes in our opinion in a sufficient way several stylized facts of high-frequency data - namely the discreteness of prices, the bid-ask bounce and time-varying bid-ask spread and the form of autocorrelations and partial autocorrelations of real log-returns (see Figure~\ref{fig:SimulatedvsRealData}). At the same time this model is more parsimonious than other models.

\medskip

\begin{figure}
\centering
\includegraphics[width=430pt,keepaspectratio]{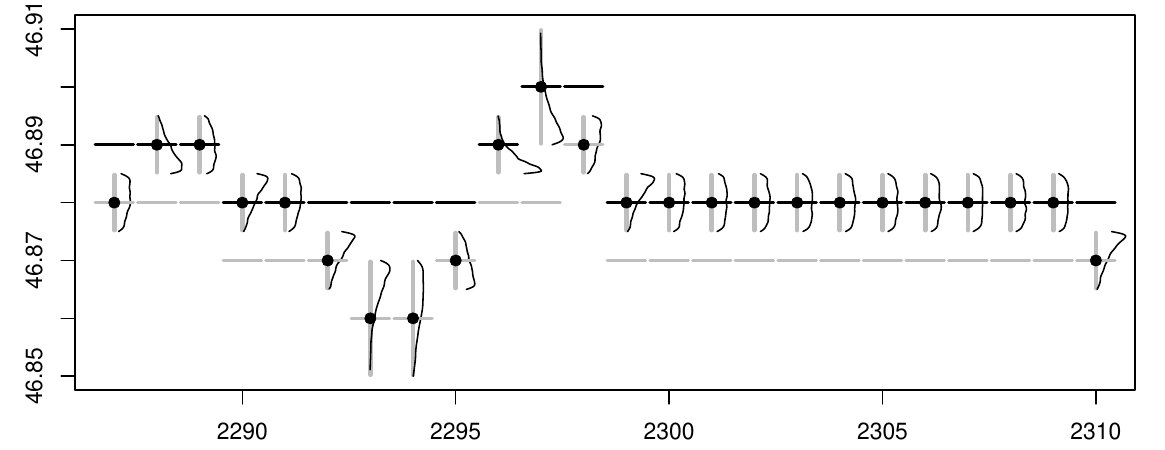}
\caption{\footnotesize
A real data example of estimated filtering distributions based on the market
microstructure noise model with deterministic rounding for the case when market maker quotes are available
in addition to the transaction data. The details are provided in
Section~\ref{ch:msn:sim:realdata}. The plot shows some transaction prices
(circles) along with kernel density estimates of the filtering distributions
of the efficient prices (black lines) based on the particles produced by the
particle filter. The gray vertical lines indicate the assumed  support of the
filtering distributions. The bid and ask market maker quotes are displayed by
gray and black horizontal lines, respectively. The x-axis shows transaction
time.}
\label{fig:MSNFilteringDistr_realQuotes_univariate}
\end{figure}

\noindent \underline{Identifiability and approximation of the filtering distribution:} \,
(\ref{basic:pf}) and (\ref{modelass1}) imply that the joint filtering distribution
$p(\mathbf{x}_{t_{1:j}}|\mathbf{y}_{t_{1:j}})$ is uniquely defined under the model assumptions.
$A_{t_j}$ and $\log A_{t_j}$ are the supports of the filtering distribution of the efficient price
$p\big(\exp(x_{t_j})\big| y_{t_{1:j}} \big)$ and the efficient log-price $p(x_{t_j}| y_{t_{1:j}})$,
respectively. It will be shown in Section~\ref{ch:msn:ssmEMalg:pf} that the filtering distributions
can be approximated through a particle filter. A real data example is given in Figure
\ref{fig:MSNFilteringDistr_realQuotes_univariate}. It shows the supports $A_{t_j}$ (gray vertical
lines) and kernel density estimates of the filtering distributions of the efficient prices (black
lines) which are computed based on the output of the particle filter. In this example, market maker
quotes are available (see Example 3 below). The details of this example are provided in
Section~\ref{ch:msn:sim:realdata}.

\medskip

For completeness we give the state equation again.

\medskip

\noindent \textbf{Model assumption 2 (state equation / efficient price model):}

\noindent The unobserved efficient price is given by $\exp(x_{t_j})$ with
\begin{equation}\label{distributional_state-equation}
p\big(x_{t_j}\big| x_{t_{j-1}} \big) =
\mathcal{N} \big(x_{t_j} \big| x_{t_{j-1}},
\sigma_{t_j}^2\big).
\end{equation}
It is assumed that $\sigma_{t_j}^2 = \sigma^{2}(t_j)$ with a function $\sigma^{2}(\cdot)$  which is either constant or smooth in time (say Lipschitz-continuous).

\bigskip

The smoothness assumption does not need to be specified any further because we do not use it
formally. However, without this assumption the estimation procedure developed in
Section~\ref{ch:msn:ssmEMalg:em} would not make sense. Any proof of consistency of the estimates of this paper would require in addition some type of in-fill asymptotics.

\medskip

\noindent \underline{Modeling microstructure noise via the specification of $A_{t_j}$:} In order to
carry out the particle filter and the volatility estimate described later we have to specify for
the observation $y_{t_j}$ at hand the set $A_{t_j}$ i.e. the set of the possible efficient prices.
This specification is an important modeling step. We now give examples.

\medskip

\noindent \textbf{Example 1 (simple deterministic and stochastic rounding):}\\
(i) The simplest example is the rounding of $\exp(x_{t_j})$ to the nearest integer (say cent) --
i.e. $y_{t_j} = \text{round} (\exp(x_{t_j}))$. In this case $A_{t_j} = [y_{t_j} - 0.5 , y_{t_j} +
0.5 )$ and  $p\big(y_{t_j}\big| y_{t_{1:j-1}}, \exp(x_{t_j})\big) =
\mathbf{1}_{A_{t_j}}\big(\exp(x_{t_j})\big)\, \mathbf{1}_{\mathbb{N}} (y_{t_j})$.\\[3pt]
(ii) A simple stochastic example is where we choose for $\exp(x_{t_j}) \in (n,n+1)$ the values
$y_{t_j} = n$ and $y_{t_j} = n+1$ each  with
probability 1/2. In that case $A_{t_j} = (y_{t_j} - 1 , y_{t_j} + 1 )$ and\\
$p\big(y_{t_j}\big| y_{t_{1:j-1}}, \exp(x_{t_j})\big) = \frac {1} {2} \,
\mathbf{1}_{A_{t_j}}\big(\exp(x_{t_j})\big)\, \mathbf{1}_{\mathbb{N}} (y_{t_j})$ for almost all $x_{t_j}$. It seems natural to set
$y_{t_j} = n\,$ for $\,\exp(x_{t_j})=n\,$ but with order book data as in Example 2 below this
choice is no longer natural.
\bigskip

We now give some examples where we model the bid-ask spread of financial transaction data. In these
cases $A_{t_j}$ depends on past observations and/or exogenous data.

\medskip

\noindent \textbf{Example 2 (order book data):}\\
Let's assume that at each transaction time $t_j$ the exchange provides a limit order book with bid
and ask levels given by $\alpha_{t_j}^k$ $(k=1,2,\ldots,K)$ and $\beta_{t_j}^{\ell}$
$(\ell=1,2,\ldots,L)$ respectively (these are the levels where contract offers are really
available). The order book levels satisfy $\alpha_{t_j}^K < \ldots < \alpha_{t_j}^2 <
\alpha_{t_j}^1 < \beta_{t_j}^1 < \beta_{t_j}^2 < \ldots < \beta_{t_j}^L$ and we denote
\[
\mathcal{M}_{t_j \mbox{-}} := \{\alpha_{t_j}^K, \ldots, \alpha_{t_j}^2, \alpha_{t_j}^1,
\beta_{t_j}^1, \beta_{t_j}^2, \ldots, \beta_{t_j}^L \}.
\]
$\mathcal{M}_{t_j \mbox{-}}$ represents the state of the order book immediately before the
transaction at time $t_j$ occurs. $\mathcal{M}_{t_j \mbox{-}}$ depends in an unknown way on the
past observations $y_{t_{1:j-1}}$ and exogenous information. Clearly $y_{t_j} \in \mathcal{M}_{t_j
\mbox{-}}\,$. We now set, corresponding to the deterministic case (i) in Example 1
\begin{equation} \label{ob_deterministic-round}
A_{t_j} := \{z \in \mathbb{R}:
\text{argmin}_{\gamma \in \mathcal{M}_{t_j \mbox{-}}} |z-\gamma| = y_{t_j} \}
\end{equation}
or equivalently
\begin{equation*} \label{}
g_{t_j;y_{t_{1:j-1}}}\!(z) : = \text{argmin}_{\gamma \in \mathcal{M}_{t_j \mbox{-}}}
|z-\gamma|\,.
\end{equation*}
Thus the transaction price at time $t_j$ is that price from $\mathcal{M}_{t_j \mbox{-}}$ with the
smallest Euclidean distance to the efficient price. This means that the efficient price at time
$t_j$ is assumed to be closer to the observed price $y_{t_j}$ than to any other order book level.
Of course, this cannot be guaranteed and it seems to be more realistic to choose for (say) $\gamma
\in (\alpha_{t_j}^1, \beta_{t_j}^1)$   $\;y_{t_j}=\alpha_{t_j}^1$ and $ y_{t_j}=\beta_{t_j}^1$ each
with probability $1/2$ - i.e. a trade is made with probability $1/2$ on the bid and on the ask
side. This corresponds to the stochastic case (ii) from Example 1 leading to the definition
$A_{t_j} := \big($largest level from $\mathcal{M}_{t_j \mbox{-}}$ below $y_{t_j},$ smallest level
from $\mathcal{M}_{t_j \mbox{-}}$ above $y_{t_j}\big)$. Figure~\ref{fig:SimulatedvsRealData}
indicates that the model with $A_{t_j}$ as in (\ref{ob_deterministic-round}) (deterministic
rounding)  better captures the stylized facts of real transaction data. The explanation may be
that in the case of a liquid order book often several trades are executed at the same level and
therefore the first model gives a better fit. The situation may be different if the stock is less
heavily traded - however we have not investigated that.

In the present situation we could better write instead of (\ref{modelass1})
\begin{equation*}
\label{modelass1b}
p\big(y_{t_j}\big| \mathcal{M}_{t_j \mbox{-}}, \exp(x_{t_j})\big) \propto
\mathbf{1}_{A_{t_j}}\big(\exp(x_{t_j})\big)  \, \mathbf{1}_{\mathcal{Y}} (y_{t_j}) \; \mbox{ a.s.}
\end{equation*}
where $\mathcal{M}_{t_j \mbox{-}}$ contains implicitly the relevant information from
$y_{t_{1:j-1}}$.

If the volume of the trade at time $t_j$ is so large that it is executed on several levels of the
order book then $y_{t_j}$ should be set equal to the largest ask level (smallest bid level) and all
lower levels should be deleted before determining $A_{t_j}$.

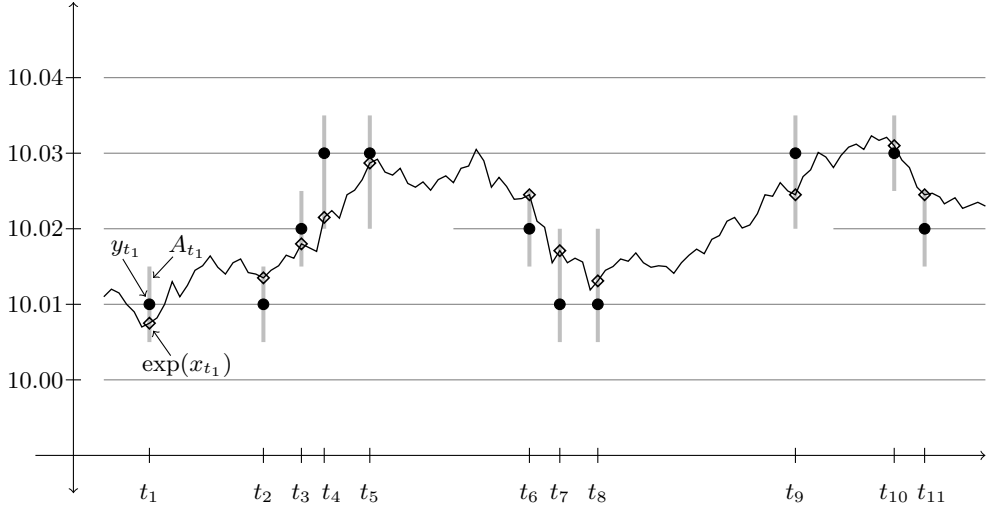
\begin{figure}[t]
\centering

\begin{tikzpicture}

\draw [lightgray, line width=1.5pt] (1,2.5) -- (1,1.5); \draw [lightgray, line width=1.5pt]
(2.5,2.5) -- (2.5,1.5); \draw [lightgray, line width=1.5pt] (3,2.5) -- (3,3.5); \draw [lightgray,
line width=1.5pt] (3.3,3) -- (3.3,4.5); \draw [lightgray, line width=1.5pt] (3.9,3) -- (3.9,4.5);
\draw [lightgray, line width=1.5pt] (6,2.5) -- (6,3.5); \draw [lightgray, line width=1.5pt] (6.4,3)
-- (6.4,1.5); \draw [lightgray, line width=1.5pt] (6.9,3) -- (6.9,1.5); \draw [lightgray, line
width=1.5pt] (9.5,3) -- (9.5,4.5); \draw [lightgray, line width=1.5pt] (10.8,3.5) -- (10.8,4.5);
\draw [lightgray, line width=1.5pt] (11.2,2.5) -- (11.2,3.5);

\draw [black, <-] (12,0) -- (-0.5,0); \draw [black, <->] (0,6) -- (0,-0.5);

\draw [gray] (12,1) -- (0.4,1); \draw [gray] (12,2) -- (0.4,2); \draw [gray] (3,3) -- (0.4,3);
\draw [gray] (6,3) -- (5,3); \draw [gray] (12,3) -- (10,3); \draw [gray] (12,4) -- (0.4,4); \draw
[gray] (12,5) -- (0.4,5);

\filldraw [black] (1,2) circle (2pt) (2.5,2) circle (2pt) (3,3) circle (2pt) (3.3,4) circle (2pt)
(3.9,4) circle (2pt) (6,3) circle (2pt) (6.4,2) circle (2pt) (6.9,2) circle (2pt) (9.5,4) circle
(2pt) (10.8,4) circle (2pt) (11.2,3) circle (2pt);

\draw[line width=0.60pt] (1+0,1.75+0.075) -- (1+0.075,1.75+0) -- (1+0,1.75-0.075) --
(1-0.075,1.75+0) -- cycle; \draw[line width=0.60pt] (2.5+0,2.35+0.075) -- (2.5+0.075,2.35+0) --
(2.5+0,2.35-0.075) -- (2.5-0.075,2.35+0) -- cycle; \draw[line width=0.60pt] (3+0,2.8+0.075) --
(3+0.075,2.8+0) -- (3+0,2.8-0.075) -- (3-0.075,2.8+0) -- cycle; \draw[line width=0.60pt]
(3.3+0,3.15+0.075) -- (3.3+0.075,3.15+0) -- (3.3+0,3.15-0.075) -- (3.3-0.075,3.15+0) -- cycle;
\draw[line width=0.60pt] (3.9+0,3.87+0.075) -- (3.9+0.075,3.87+0) -- (3.9+0,3.87-0.075) --
(3.9-0.075,3.87+0) -- cycle; \draw[line width=0.60pt] (6+0,3.45+0.075) -- (6+0.075,3.45+0) --
(6+0,3.45-0.075) -- (6-0.075,3.45+0) -- cycle; \draw[line width=0.60pt] (6.4+0,2.71+0.075) --
(6.4+0.075,2.71+0) -- (6.4+0,2.71-0.075) -- (6.4-0.075,2.71+0) -- cycle; \draw[line width=0.60pt]
(6.9+0,2.31+0.075) -- (6.9+0.075,2.31+0) -- (6.9+0,2.31-0.075) -- (6.9-0.075,2.31+0) -- cycle;
\draw[line width=0.60pt] (9.5+0,3.45+0.075) -- (9.5+0.075,3.45+0) -- (9.5+0,3.45-0.075) --
(9.5-0.075,3.45+0) -- cycle; \draw[line width=0.60pt] (10.8+0,4.10+0.075) -- (10.8+0.075,4.10+0) --
(10.8+0,4.10-0.075) -- (10.8-0.075,4.10+0) -- cycle; \draw[line width=0.60pt] (11.2+0,3.45+0.075)
-- (11.2+0.075,3.45+0) -- (11.2+0,3.45-0.075) -- (11.2-0.075,3.45+0) -- cycle;

\draw [black] (1,-0.1) -- (1,0.1); \draw [black] (2.5,-0.1) -- (2.5,0.1); \draw [black] (3,-0.1) --
(3,0.1); \draw [black] (3.3,-0.1) -- (3.3,0.1); \draw [black] (3.9,-0.1) -- (3.9,0.1); \draw
[black] (6,-0.1) -- (6,0.1); \draw [black] (6.4,-0.1) -- (6.4,0.1); \draw [black] (6.9,-0.1) --
(6.9,0.1); \draw [black] (9.5,-0.1) -- (9.5,0.1); \draw [black] (10.8,-0.1) -- (10.8,0.1); \draw
[black] (11.2,-0.1) -- (11.2,0.1);

\node at (1,-0.5) {\footnotesize $t_1$}; \node at (2.5,-0.5) {\footnotesize $t_2$}; \node at
(3,-0.5) {\footnotesize $t_3$}; \node at (3.4,-0.5) {\footnotesize $t_4$}; \node at (3.9,-0.5)
{\footnotesize $t_5$}; \node at (6,-0.5) {\footnotesize $t_6$}; \node at (6.4,-0.5) {\footnotesize
$t_7$}; \node at (6.9,-0.5) {\footnotesize $t_8$}; \node at (9.5,-0.5) {\footnotesize $t_9$}; \node
at (10.8,-0.5) {\footnotesize $t_{10}$}; \node at (11.3,-0.5) {\footnotesize $t_{11}$};

\draw [black] (-0.1,1) -- (0.1,1); \draw [black] (-0.1,2) -- (0.1,2); \draw [black] (-0.1,3) --
(0.1,3); \draw [black] (-0.1,4) -- (0.1,4); \draw [black] (-0.1,5) -- (0.1,5);

\node at (-0.5,1) {\footnotesize $10.00$}; \node at (-0.5,2) {\footnotesize $10.01$}; \node at
(-0.5,3) {\footnotesize $10.02$}; \node at (-0.5,4) {\footnotesize $10.03$}; \node at (-0.5,5)
{\footnotesize $10.04$};

\node at (0.7,2.73) {\footnotesize   $y_{t_1}$}; \draw [black, ->] (0.65,2.6) -- (0.95,2.1); \node
at (1.5,1.2) {\footnotesize $\exp(x_{t_1})$}; \draw [black, ->] (1.28,1.32) -- (1.05,1.65); \node
at (1.50,2.76) {\footnotesize $A_{t_1}$}; \draw [black, ->] (1.31,2.63) -- (1.05,2.35);

\draw [black, line width=0.5pt] (0.4,2.1) -- (0.5,2.2) -- (0.6,2.15) -- (0.7,2) -- (0.8,1.9) --
(0.9,1.7)  -- (1,1.75)

-- (1.1,1.82) -- (1.2,2.0) -- (1.3,2.3) -- (1.4,2.1) -- (1.5,2.25) -- (1.6,2.45) -- (1.7,2.51) --
(1.8,2.64) -- (1.9,2.49)  -- (2,2.4)

-- (2.1,2.55) -- (2.2,2.6) -- (2.3,2.42) -- (2.4,2.4) -- (2.5,2.35) -- (2.6,2.45) -- (2.7,2.51) --
(2.8,2.65) -- (2.9,2.61)  -- (3,2.8)

-- (3.1,2.75) -- (3.2,2.7) -- (3.3,3.15) -- (3.4,3.24) -- (3.5,3.14) -- (3.6,3.45) -- (3.7,3.51) --
(3.8,3.65) -- (3.9,3.87)  -- (4,3.92)

-- (4.1,3.75) -- (4.2,3.71) -- (4.3,3.8) -- (4.4,3.6) -- (4.5,3.55) -- (4.6,3.62) -- (4.7,3.51) --
(4.8,3.65) -- (4.9,3.70)  -- (5,3.61)

-- (5.1,3.8) -- (5.2,3.83) -- (5.3,4.05) -- (5.4,3.9) -- (5.5,3.55) -- (5.6,3.68) -- (5.7,3.56) --
(5.8,3.39) -- (5.9,3.40)  -- (6,3.45)

-- (6.1,3.1) -- (6.2,3.02) -- (6.3,2.55) -- (6.4,2.71) -- (6.5,2.55) -- (6.6,2.61) -- (6.7,2.56) --
(6.8,2.19) -- (6.9,2.31)  -- (7,2.45)

-- (7.1,2.5) -- (7.2,2.6) -- (7.3,2.57) -- (7.4,2.68) -- (7.5,2.55) -- (7.6,2.49) -- (7.7,2.51) --
(7.8,2.5) -- (7.9,2.41)  -- (8,2.55)

-- (8.1,2.65) -- (8.2,2.73) -- (8.3,2.67) -- (8.4,2.86) -- (8.5,2.91) -- (8.6,3.1) -- (8.7,3.15) --
(8.8,3.01) -- (8.9,3.05)  -- (9,3.2)

-- (9.1,3.45) -- (9.2,3.43) -- (9.3,3.61) -- (9.4,3.5) -- (9.5,3.45) -- (9.6,3.69) -- (9.7,3.78) --
(9.8,4.01) -- (9.9,3.95)  -- (10,3.81)

-- (10.1,3.97) -- (10.2,4.08) -- (10.3,4.12) -- (10.4,4.05) -- (10.5,4.23) -- (10.6,4.17) --
(10.7,4.21) -- (10.8,4.10) -- (10.9,3.91)  -- (11,3.81)

-- (11.1,3.55) -- (11.2,3.45) -- (11.3,3.47) -- (11.4,3.42) -- (11.46,3.33) -- (11.6,3.41) --
(11.7,3.27) -- (11.8,3.31) -- (11.9,3.35)  -- (12,3.30);

\end{tikzpicture}
\caption{\footnotesize  An example of the market microstructure noise model with deterministic
rounding for the case when order book data are available. The figure shows the transaction prices
(circles), the (in practice unknown) efficient prices in transaction time (diamonds), the latent
efficient price process in clock time (black line), the order book levels (gray horizontal lines),
and the supports of the filtering distributions of the efficient prices (gray vertical lines).}
\label{fig:marketmicrostucturemodel}
\end{figure}

An example of this market microstructure model is visualized in Figure
\ref{fig:marketmicrostucturemodel}. The intervals $A_{t_j}$ are denoted by thick vertical lines.
Note, that these are also the supports of the filtering distributions. Larger intervals $A_{t_j}$
are usually due to a larger bid-ask spread.

\bigskip

\noindent \textbf{Example 3 (market maker quotes):}\\
In case where market maker quotes are available instead of order book data, we only have a single
bid and a single ask level $\alpha_{t_j}$ and $\beta_{t_j}$, respectively, which satisfy
$\alpha_{t_j} < \beta_{t_j}$. That is, $y_{t_j}$ is either equal to $\alpha_{t_j}$ or equal to
$\beta_{t_j}$. Corresponding to deterministic rounding as in Example 1 (i) we set
\begin{equation} \label{AtEx3}
A_{t_j} = [y_{t_j} -  \Delta_{t_j},
y_{t_j} + \Delta_{t_j})
\end{equation}
where $ \Delta_{t_j} := 0.5\, (\beta_{t_j} - \alpha_{t_j})$. The choice $ \Delta_{t_j} :=  (\beta_{t_j} - \alpha_{t_j})$ corresponds to stochastic rounding as in Example 1 (ii).

From a certain point of view this choice of $A_{t_j}$ seems to be not adequate and one is tempted
to chose
\begin{equation*}
A_{t_j} = \left\{
\begin{array}{ll}
\big(\!-\!\infty \, , \, \alpha_{t_j} \!+   \Delta_{t_j}\big), &  \mbox{if } y_{t_j} = \alpha_{t_j}\\[6pt]
\big[\beta_{t_j} \!- \Delta_{t_j}\, , \, \infty\big), &  \mbox{if } y_{t_j} = \beta_{t_j}
\end{array} \right..
\end{equation*}

\noindent To understand why this is not a proper choice one needs to look in more detail at the
behavior of the market maker. Of course the market maker has more (invisible) levels which are
automatically executed at the same time if the efficient price makes larger jumps. Furthermore, the
market maker has additional information on the efficient price (say from trades of correlated
securities) and may have already adjusted his levels towards the efficient price. This last fact
violates our model assumptions (in that the function $g_{t_j}$ not only depends on past values
$y_{t_{1:j-1}}$ and exogenous information but also somehow on $\exp(x_{t_j})$) but in particular in
this situation our model with the above choice of $A_{t_j}$ seems to be a reasonable parsimonious
model.

This example also demonstrates the advantage of the fact that we just have to specify the inverse
image $A_{t_j}$ for the $y_{t_j}$ at hand.

\medskip

\noindent \textbf{Example 4 (transaction data only):}\\
We now consider the situation where no order book data or market maker quotes are available. In
this case we try to estimate the order book levels from the data and use afterwards the $A_{t_j}$
from Example 2. More precisely we estimate half the bid-ask spread at time $t_j$ by
\begin{equation} \label{AtEx4a}
 \Delta_{t_j} =
 \begin{cases}
 0.5 \, |y_{t_j}-y_{t_{j-1}}| & \text{if} \;  y_{t_j} \not= y_{t_j-1}, \\
 \Delta_{t_{j-1}} & \text{else}
 \end{cases}
\end{equation}
and set
\begin{equation} \label{AtEx4}
A_{t_j} = [y_{t_j} -\Delta_{t_j}, y_{t_j} +\Delta_{t_j})
\end{equation}
(for stochastic rounding we delete the factor $0.5$ above). Surprisingly, this specification does not belong to a deterministic but to a stochastic mapping
$g_{t_j}$ (it is not difficult to see that the same $x_{t_j}$ lies in $A_{t_j}$ constructed from
different $y_{t_j}$ - hence $g_{t_j}$ must be stochastic). The mapping $g_{t_j}$ becomes
deterministic (conditionally on $y_{t_{1:j-1}}$) if one replaces $\Delta_{t_j}$ by
$\Delta_{t_{j-1}}$. However $\Delta_{t_{j-1}}$ obviously is much worse than $\Delta_{t_j}$ as an
estimate of the bid-ask spread at time $t_j$ - so from a practical point of view the above
specification is to be preferred.

\bigskip

We finally demonstrate in a simulation example that the model reproduces the autocorrelations and
partial autocorrelations of real log-returns. In addition we compare the deterministic  and the
stochastic rounding from Example 1. In Figure~\ref{fig:SimulatedvsRealData} transaction data of
Citigroup are compared with data simulated from our model with the two different rounding schemes
from Example 1(i) and (ii), respectively. The figure shows the simulated efficient and the observed
prices. The efficient log-prices were generated according to (\ref{rounding:stateequation}) such
that the observations have approximately the same volatility as the Citigroup data.
The important point is that the market microstructure noise model
with deterministic rounding automatically introduces autocorrelations and partial autocorrelations
of the log-returns which are similar to those of the real Citigroup data while the results with stochastic rounding are worse. Another indicator for the superiority of deterministic vs. stochastic rounding are the results of Section~\ref{ch:msn:sim:realdata} (see the paragraph ``Transaction time volatility estimation'' and Figure~\ref{fig:RealData_allEstimatorsStartValue}). In addition, the model covers bid-ask bounces, time-varying bid-ask spreads and price discreteness.

Although there is some evidence to use deterministic rounding, all methods of this paper can also be applied with stochastic models: The stochastic rounding discussed above can also be used in combination with order book data (Example 2), market maker quotes (Example 3) and pure transaction data (Example 4). In particular the corresponding volatility estimator with stochastic rounding is included in Figure~\ref{fig:RealData_allEstimatorsStartValue}. Other types of stochastic rounding such as $Y_{t_j} = \text{round}\big(\exp(X_{t_j} + U_{t_j})\big)$ are discussed at the end of Section~\ref{ch:msn:ssmEMalg:pf}.

\section{On-Line Estimation of Spot Volatility} \label{ch:msn:ssmEMalg}

We now present on-line algorithms for the estimation of the spot volatility.
Because all results also hold in the multivariate case \underline{with
synchronous trading times} we formulate this section for multivariate
security prices. We are aware of the fact that the main challenge in the
multivariate case are non-synchronous trading times. The presented results
are, however, the basis for future work on non-synchronous trading.

We therefore consider in this section the estimation of the covariance matrix $\Sigma_{t_j}$ which
gives the volatilities of the individual efficient log-price processes $\mathbf{X}_t \!=\!
(X_{t,1}, \ldots, X_{t,S})'$ as well as their cross-volatilities. Thus for $S=1$ we have  $\Sigma_{t_{j}} = \sigma_{t_{j}}^2$. The multivariate version of
the nonlinear state-space model (\ref{modelass1}) and (\ref{distributional_state-equation}) is
given by
\begin{eqnarray}
\label{roundssm:synchObsEquation}
p\big(\mathbf{y}_{t_j}\big| \mathbf{y}_{t_{1:j-1}}, \exp(\mathbf{x}_{t_j})\big)
& \propto & \mathbf{1}_{\mathbf{A}_{t_j}}\big(\exp(\mathbf{x}_{t_j})\big)  \, \mathbf{1}_{\mathcal{Y}} (\mathbf{y}_{t_j}) \quad \mbox{a.s.},\\
\label{roundssm:synchStateEquation}
p\big(\mathbf{x}_{t_j}\big| \mathbf{x}_{t_{j-1}} \big) &=&
\mathcal{N}\big(\mathbf{x}_{t_j} \big| \mathbf{x}_{t_{j-1}};
\Sigma_{t_{j}}\big),
\end{eqnarray}
where $\mathbf{X}_t \!=\! (X_{t,1}, \ldots, X_{t,S})'$, $\mathbf{Z}_{t_j} \!\sim\!
\mathcal{N}(\mathbf{0}, \Sigma_{t_{j}})$, and the {\rm exp}-function is applied componentwise. The set $\mathbf{A}_{t_{j}}$ usually is of the form
$\mathbf{A}_{t_{j}} = A_{t_{j},1} \times \dots \times A_{t_{j},S}$. We assume that model assumption
1 holds for all components. For simplicity we assume as an initial condition that given
$Y_{t_{1},s} = \mathbf{y}_{t_1,s}$ the efficient prices $\exp(X_{t_{1},s})$ are uniformly
distributed on $A_{t_1,s}$.

We remark that (\ref{roundssm:synchObsEquation}) and
(\ref{roundssm:synchStateEquation}) constitute a slightly generalized
state-space model because the observations $\mathbf{Y}_{t_j}$ are not
conditional independent of $\mathbf{Y}_{t_{1:j-1}}$ given $\mathbf{X}_{t_j}$ as
in standard state-space models. However, this is a standard extension which
does not cause any difficulty for estimation.

Our objective is the estimation of the covariance matrix
$\Sigma_{t_j}$ based on the observed prices $\mathbf{y}_{t_{1:j}}$.
Because of the nonlinear market microstructure noise this is difficult.
It is well known that crude estimators that ignore the noise lead to severely
biased estimates (see, for instance, Voev and Lunde 2007).
The idea of our estimation procedure is to approximate the
conditional distribution of the efficient log-prices $\mathbf{X}_{t_j}$ given
all observed transaction prices $\mathbf{y}_{t_{1:j}}$ up to time $t_j$ by an efficient
particle filter. Based on this approximation a localized EM-type
algorithm is used to construct an estimator of $\Sigma_{t_j}$.

We mention that, for such a state-space model, particle filters have been used before (Andrieu and Doucet 2002).
An alternative to particle filters would be to use MCMC methods together with an EM algorithm (Manrique and Shepard 1998).
A related model with rounding noise is discussed by Hasbrouck (1999, 2004) who also used MCMC methods for estimation.
In comparison to the existing literature we provide a more general model for microstructure noise and focus on localized estimation. Owens and Steigerwald (2006) have used a Kalman filter in a linear microstructure noise model to derive volatility estimates based on weighted observations. Their method can be modified for on-line estimation of spot volatility.

\subsection{An Efficient Particle Filter} \label{ch:msn:ssmEMalg:pf}

Particle filters are sequential Monte Carlo methods (Doucet et al. 2001) that
approximate the posterior distributions $p(\mathbf{x}_{t_{1:j}}|\mathbf{y}_{t_{1:j}})$ with clouds of particles
$\{\mathbf{x}_{t_{1:j}}^i,  \omega_{t_{j}}^i \}_{i=1}^N$. A particle consists
of a sample $\mathbf{x}_{t_{1:j}}^i$ and an associated weight
$\omega_{t_{j}}^i$.
The particle approximation of the target distribution is given by
\[
p(\mathbf{x}_{t_{1:j}}|\mathbf{y}_{t_{1:j}}) \approx \sum_{i=1}^N
\omega_{t_{j}}^i \delta_{\mathbf{x}_{t_{1:j}}^i}(\mathbf{x}_{t_{1:j}}),
\]
with $\delta$ being the Dirac delta function.
A particle filter generates particles sequentially in time making use of the
relation
\begin{equation}\label{basic:pf}
p(\mathbf{x}_{t_{1:j}}|\mathbf{y}_{t_{1:j}}) =
\frac {p(\mathbf{y}_{t_j},\mathbf{x}_{t_{1:j}}|\mathbf{y}_{t_{1:j-1}})}
{p(\mathbf{y}_{t_j}|\mathbf{y}_{t_{1:j-1}})} =
\frac{p(\mathbf{y}_{t_j}|\mathbf{y}_{t_{1:j-1}}, \mathbf{x}_{t_j}) \,
p(\mathbf{x}_{t_j}|\mathbf{x}_{t_{j-1}})}
{p(\mathbf{y}_{t_j}|\mathbf{y}_{t_{1:j-1}})} \;
p(\mathbf{x}_{t_{1:j-1}}|\mathbf{y}_{t_{1:j-1}})
\end{equation}
and a general sampling technique known as importance sampling. Importance
sampling is necessary because direct sampling from (\ref{basic:pf}) is
not feasible.
In standard state-space models $p(\mathbf{y}_{t_j}|\mathbf{y}_{t_{1:j-1}},
\mathbf{x}_{t_j})$ further simplifies to
$p(\mathbf{y}_{t_j}|\mathbf{x}_{t_j})$. As a result of the violated conditional
independence property mentioned earlier, this is not the case here.

In each iteration of the particle filter samples are drawn from an importance
sampling distribution called proposal. Subsequently, the
samples are weighted such that they approximate the target distribution. The
choice of the proposal is crucial for the efficiency of the filter. In our framework
it is possible to sample from the proposal
$p(\mathbf{x}_{t_{j}}|\mathbf{y}_{t_{1:j}}, \mathbf{x}_{t_{j-1}})$ which is the
optimal proposal in the sense that it minimizes the variance of the importance
sampling weights (Doucet et al. 2000). The algorithm can be stated as
follows:
Assume that weighted particles $\{\mathbf{x}_{t_{1:j-1}}^i,
\omega_{t_{j-1}}^i \}_{i=1}^N$ approximating $p(\mathbf{x}_{t_{1:j-1}}|\mathbf{y}_{t_{1:j-1}})$
are given; then
\begin{itemize}
\baselineskip1.1em
\itemsep-0.03cm
\item For $i=1, \ldots, N$:
\begin{itemize}
\baselineskip1.5em
\itemsep-0.03cm
  \item Sample from the optimal proposal: $\,\mathbf{x}_{t_{j}}^i \sim
  p(\mathbf{x}_{t_j}|\mathbf{y}_{t_{1:j}}, \mathbf{x}_{t_{j-1}}^i)$.
  \item Compute importance weights
\[\breve{\omega}_{t_j}^i \propto
  \omega_{t_{j-1}}^i
  \frac{p(\mathbf{y}_{t_j}|\mathbf{y}_{t_{1:j-1}}, \mathbf{x}_{t_j}^i)\,
  p(\mathbf{x}_{t_j}^i|\mathbf{x}_{t_{j-1}}^i) }
  {p(\mathbf{x}_{t_j}^i|\mathbf{y}_{t_{1:j}}, \mathbf{x}_{t_{j-1}}^i)} =
  \omega_{t_{j-1}}^i \,
  p(\mathbf{y}_{t_j}|\mathbf{y}_{t_{1:j-1}}, \mathbf{x}_{t_{j-1}}^i).\]
  \end{itemize}
  \item For $i=1, \ldots, N$:
\begin{itemize}
  \item Normalize importance weights $\omega_{t_j}^i = \breve{\omega}_{t_j}^i/(
\sum_{k=1}^N \breve{\omega}_{t_j}^k)$.
\end{itemize}
  \item Obtain particles $\{\mathbf{x}_{t_{1:j}}^i,  \omega_{t_{j}}^i
  \}_{i=1}^N$ which approximate
  $p(\mathbf{x}_{t_{1:j}}|\mathbf{y}_{t_{1:j}})$.
\end{itemize}
It is well-known that this algorithm suffers from weight degeneracy which
means that after some iterations only few particles will have significant
weight.
This issue can be resolved by introducing a resampling step that maps
the particle system $\{\mathbf{x}_{t_{1:j}}^i, \omega_{t_{j}}^i \}_{i=1}^N$
onto an equally weighted particle system $\{\mathbf{x}_{t_{1:j}}^i, 1/N
\}_{i=1}^N$. Because resampling is time-consuming, it is carried out only if
the effective sample size
\begin{equation*}
\text{ESS}\big(\{\omega_{t_{j}}^i\}_{i=1}^N\big) = \frac{1}{\sum_{i=1}^N (\omega_{t_{j}}^i)^2}
\end{equation*}
is below some threshold (Kong et al. 1994). Other resampling schemes are discussed in Douc et al.
(2005).

To apply this particle filter to the state-space model given by
(\ref{roundssm:synchObsEquation}) and (\ref{roundssm:synchStateEquation}) it is
necessary to specify the optimal proposal and the computation of the importance
weights. The following result shows that both take a very simple form.

\begin{prop}\label{msn:prop1}
The optimal proposal is a truncated multivariate normal distribution given by
\begin{equation*} \label{OptimalProposal}
p(\mathbf{x}_{t_{j}}|\mathbf{y}_{t_{1:j}}, \mathbf{x}_{t_{j-1}}) \propto
\mathcal{N}(\mathbf{x}_{t_{j}}| \mathbf{x}_{t_{j-1}};
\Sigma_{t_{j}})\big|_{\log \mathbf{A}_{t_{j}}}
\end{equation*}
with $\log \mathbf{A}_{t_{j}} = \log A_{t_{j},1} \times \dots \times
\log A_{t_{j},S}$
and the importance weights can be computed through
\begin{equation}\label{IScomputationEq}
\breve{\omega}_{t_{j}}^i \propto \omega_{t_{j-1}}^i
\int_{\log \mathbf{A}_{t_{j}}} \mathcal{N}(\mathbf{x}_{t_{j}}|
\mathbf{x}_{t_{j-1}}^i; \Sigma_{t_{j}}) \, d\mathbf{x}_{t_{j}}.
\end{equation}
\end{prop}

\bigskip

\noindent \textbf{Proof:}  The proof is straightforward.

\medskip

\noindent \textbf{Remark (rounding with additive noise):} Alternative stochastic models with
rounding are
\begin{equation}\label{alternat:msnmodel}
Y_{t_j} = \text{round} \big(\exp(X_{t_j}) + U_{t_j}\big) \qquad \mbox{or}
\qquad Y_{t_j} = \text{round}\big(\exp(X_{t_j} + U_{t_j})\big)
\end{equation}
with i.i.d. Gaussian $U_{t_j}$. Let $\mathbf{A}_{t_j} = [y_{t_j} - 0.5 , y_{t_j}
+ 0.5 )$. For example for the second model we consider the corresponding state space model with state variable
$\tilde{X}_{t_j} = \big(X_{t_j}, U_{t_j}\big)'$. Then the optimal proposal
satisfies
\begin{align*}
p(\tilde{x}_{t_{j}}|y_{t_{1:j}}, \tilde{x}_{t_{j-1}}) & \propto
p(y_{t_j}|y_{t_{1:j-1}}, \tilde{x}_{t_j})\,
  p(\tilde{x}_{t_j}|\tilde{x}_{t_{j-1}})\\[6pt]
& =   p(x_{t_j}|x_{t_{j-1}}) \, p(u_{t_j}) \, \mathbf{1}_{\mathbf{A}_{t_j}} \!\big(\exp(x_{t_j} + u_{t_j})\big)
\end{align*}
which can be used easily to sample the particles in the filter step. The
importance weights are given by
\begin{align*}
p(y_{t_j}|y_{t_{1:j-1}}, \tilde{x}_{t_{j-1}}^i)
& = \, p(y_{t_j}|y_{t_{1:j-1}}, x_{t_{j-1}}^i)\\
& = \int \!\!\! \int \mathbf{1}_{\log \mathbf{A}_{t_j}} \big(x_{t_j} + u_{t_j})\,
\mathcal{N}(x_{t_j}|x_{t_{j-1}}^{i};\sigma_{t_j}^{2})\,
\mathcal{N}(u_{t_j}|\, 0;\sigma_{U}^{2}) \, dx_{t_j} du_{t_j}
\end{align*}
which are however more difficult to compute (compare Section~\ref{ch:msn:impl}).

\subsection{A Sequential EM-Type Algorithm} \label{ch:msn:ssmEMalg:em}

In this section, we discuss the estimation of $\Sigma_{t_j}$ in the
time-constant and time-varying case.

A stochastic EM algorithm can be used to obtain the maximum likelihood
estimator in the time-constant case $\Sigma_{t_{j}} = \Sigma$ (Dempster et al.
1977). The EM algorithm maximizes the likelihood
$p_{\Sigma}(\mathbf{y}_{t_{1:T}})$ by iteratively carrying out an E-step and
an M-step. In the E-step, the expectation
\begin{eqnarray}
\nonumber
\mathcal{Q}(\Sigma| \hat{\Sigma}^{(m)})
&=& \mathbf{E}_{\hat{\Sigma}^{(m)}}\big[\log
p_{\Sigma}(\mathbf{X}_{t_{1:T}}, \mathbf{y}_{t_{1:T}}) | \mathbf{y}_{t_{1:T}}\big]\\
\nonumber
&=& \sum_{j=1}^T \mathbf{E}_{\hat{\Sigma}^{(m)}}\big[\log
p(\mathbf{y}_{t_{j}}| \mathbf{y}_{t_{1:j-1}}, \mathbf{X}_{t_{j}}) |
\mathbf{y}_{t_{1:T}}\big] + \mathbf{E}_{\hat{\Sigma}^{(m)}}\big[\log
p(\mathbf{X}_{t_{1}}) | \mathbf{y}_{t_{1:T}}\big]\\
\label{EM:Estep}
& & \; +
\sum_{j=2}^T \mathbf{E}_{\hat{\Sigma}^{(m)}}\big[\log
p_{\Sigma}(\mathbf{X}_{t_{j}}| \mathbf{X}_{t_{j-1}}) | \mathbf{y}_{t_{1:T}}\big]
\end{eqnarray}
needs to be approximated, where $\hat{\Sigma}^{(m)}$ is the current
estimator. Note, it is sufficient to consider the sum in (\ref{EM:Estep})
because the random variables $\log p(\mathbf{y}_{t_{j}}|
\mathbf{y}_{t_{1:j-1}}, \mathbf{X}_{t_{j}})$ and $p(\mathbf{X}_{t_{1}})$ do not
depend on $\Sigma$. In the M-step, a new parameter estimate
$\hat{\Sigma}^{(m+1)}$ is obtained by maximizing
$\mathcal{Q}(\Sigma| \hat{\Sigma}^{(m)})$. Below we show how to modify this procedure towards an on-line estimator.

If $\Sigma_{t_{j}}$ is time-varying some regularization is needed. For example
$\hat{\Sigma}_{t_j}^{(m+1)}$ can be obtained by maximizing some localized version of (\ref{EM:Estep}), e.g.
\begin{equation} \label{EM:EstepKernelLikelihood}
\mathcal{Q}_{t_{j}}(\Sigma| \hat{\Sigma}_{t_{1:T}}^{(m)}) = \frac {1} {T}
\sum_{k=j-T}^{j-2} \frac {1} {b} K\Big(\frac {k} {bT}\Big)  \,
\mathbf{E}_{\hat{\Sigma}_{t_{1:T}}^{(m)}}\big[\log p_{\Sigma} (\mathbf{X}_{t_{j-k}}|
\mathbf{X}_{t_{j-k-1}}) | \mathbf{y}_{t_{1:T}}\big]
\end{equation}
with a kernel $K(\cdot)$ and a bandwidth $b$. If the kernel is an one sided exponential kernel this can be written in recursive form as
\begin{equation}\label{seqEM1a}
\mathcal{Q}_{t_{j}}(\Sigma| \hat{\Sigma}_{t_{1:T}}) :=
\{1-\lambda_{j}\} \, \mathcal{Q}_{t_{j-1}}(\Sigma|
\hat{\Sigma}_{t_{1:T}}) + \lambda_{j} \,
\mathbf{E}_{\hat{\Sigma}_{t_{1:T}}}\big[\log p_{\Sigma}(\mathbf{X}_{t_{j}}| \mathbf{X}_{t_{j-1}}) | \mathbf{y}_{t_{1:T}}\big]
\end{equation}
with $\mathcal{Q}_{t_{2}}(\Sigma| \hat{\Sigma}_{t_{1:T}}) = \mathbf{E}_{\hat{\Sigma}_{t_{1:T}}}\big[\log
p_{\Sigma} (\mathbf{X}_{t_{2}}| \mathbf{X}_{t_{1}}) | \mathbf{y}_{t_{1:T}}\big]$ and $\lambda_j = \frac {1} {bT}$.

This procedure is not an
on-line algorithm because the conditional expectation in
(\ref{seqEM1a}) depends on all observations. Therefore,
we replace the conditioning set of variables $\{\mathbf{y}_{t_{1:T}}\}$ by
$\{\mathbf{y}_{t_{1:j}}\}$, i.e. $\mathbf{E}_{\hat{\Sigma}_{t_{1:T}}} \big[\log p_{\Sigma}(\mathbf{X}_{t_{j}}|
\mathbf{X}_{t_{j-1}}) | \mathbf{y}_{t_{1:T}}\big]$ is replaced by  $\mathbf{E}_{\hat{\Sigma}_{t_{1:j-1}}}\big[\log p_{\Sigma}(\mathbf{X}_{t_{j}}|
\mathbf{X}_{t_{j-1}}) | \mathbf{y}_{t_{1:j}}\big]$ (we need at this point an estimate for $\Sigma_{t_j}$ to apply the particle filter - see the comment at the
end of this section). This leads to the on-line algorithm
\begin{equation}\label{seqEM1f}
\mathcal{Q}_{t_{j}}(\Sigma| \hat{\Sigma}_{t_{1:j-1}}) :=
\{1-\lambda_{j}\} \, \mathcal{Q}_{t_{j-1}}(\Sigma|
\hat{\Sigma}_{t_{1:j-2}}) + \lambda_{j} \,
\mathbf{E}_{\hat{\Sigma}_{t_{1:j-1}}}\big[\log p_{\Sigma}(\mathbf{X}_{t_{j}}|
\mathbf{X}_{t_{j-1}}) | \mathbf{y}_{t_{1:j}}\big]
\end{equation}

\noindent with $\mathcal{Q}_{t_{2}}(\Sigma| \hat{\Sigma}_{t_{1}}) = \mathbf{E}_{\hat{\Sigma}_{t_{1}}}\big[\log
p_{\Sigma} (\mathbf{X}_{t_{2}}| \mathbf{X}_{t_{1}}) | \mathbf{y}_{t_{1:2}}\big]$.
$\mathcal{Q}_{t_{j}}(\Sigma| \hat{\Sigma}_{t_{1:j-1}}^{(m)})$ can be computed
with the filtering particles $\{\mathbf{x}_{t_{j-1:j}}^i,  \omega_{t_{j}}^i
\}_{i=1}^N$ from the particle filter leading to the approximation
\begin{align} \label{seqEM1fa}
\mathbf{E}_{\hat{\Sigma}_{t_{1:j-1}}} \big[\log p_{\Sigma}&(\mathbf{X}_{t_{j}}|
\mathbf{X}_{t_{j-1}}) | \mathbf{y}_{t_{1:j}}\big]\nonumber\\
& \approx \frac {1} {2}
\sum_{i=1}^N \omega_{t_j}^i \left[ S \log 2\pi + \log |\Sigma| +
\text{tr}\left\{ \Sigma^{-1}
\big(\mathbf{x}_{t_{j}}^i - \mathbf{x}_{t_{j-1}}^i\big) \big(\mathbf{x}_{t_{j}}^i -
\mathbf{x}_{t_{j-1}}^i\big)' \right\} \right].
\end{align}
The resulting estimate for $\Sigma$  can be obtained from the on-line recursion
\begin{equation} \label{seqEM1g}
\hat{\Sigma}_{t_{j}} =
\{1-\lambda_j\} \, \hat{\Sigma}_{t_{j-1}} + \lambda_j \,
\breve{\Sigma}_{t_{j}}(\omega_{t_j}) \quad \mbox{with}\quad \hat{\Sigma}_{t_{2}} = \breve{\Sigma}_{t_{2}}(\omega_{t_2})
\end{equation}
where
\begin{equation} \label{EstimateSigma4}
\breve{\Sigma}_{t_{j}}(\omega_{t_j}) := \sum_{i=1}^N \omega_{t_{j}}^i
\big(\mathbf{x}_{t_{j}}^i-\mathbf{x}_{t_{j-1}}^i\big)
\big(\mathbf{x}_{t_{j}}^i-\mathbf{x}_{t_{j-1}}^i\big)'.
\end{equation}
It can be written in closed form as
\begin{equation} \label{seqEM1gClosedForm}
\hat{\Sigma}_{t_{j}} =
\; \sum_{k=0}^{j-3}
\Big[ \prod_{\ell=0}^{k-1} (1-\lambda_{j-\ell}) \Big] \lambda_{j-k} \,
\breve{\Sigma}_{t_{j-k}}(\omega_{t_{j-k}}) + \Big[ \prod_{\ell=0}^{j-3}
(1-\lambda_{j-\ell}) \Big] \, \breve{\Sigma}_{t_{2}}(\omega_{t_2})\,.
\end{equation}
The new parameter estimate $\hat{\Sigma}_{t_{j}}$ is used afterwards to
calculate the next filtering particles and their weights
$\{\mathbf{x}_{t_{j+1}}^i,  \omega_{t_{j+1}}^i \}_{i=1}^N$ followed by the
calculation of $\hat{\Sigma}_{t_{j+1}}$ via another application of
(\ref{seqEM1g}) etc.
In contrast to the standard EM algorithm, our sequential
variant updates the covariance estimate (which in turn is used in the next
step of the particle filter) in every time step. In the ``new E-step'',
$\mathcal{Q}_{t_{j}}(\Sigma| \hat{\Sigma}_{t_{1:j-1}})$ is approximated through
(\ref{seqEM1f}) and (\ref{seqEM1fa}) using the particles $\{\mathbf{x}_{t_{j-1:j}}^i,  \omega_{t_{j}}^i
  \}_{i=1}^N$ which are generated as described in
Section~\ref{ch:msn:ssmEMalg:em}. In the ``new M-step'', the maximization of
$\hat{\mathcal{Q}}_{t_{j}}(\Sigma| \hat{\Sigma}_{t_{1:j-1}})$ gives the
on-line estimator defined in (\ref{seqEM1g}).

Note that $\breve{\Sigma}_{t_{j}}(\omega_{t_j})$ is \underline{not} an
approximation of the conditional variance $\text{Var}
\big(\mathbf{X}_{t_{j}}- \mathbf{X}_{t_{j-1}} \big| \mathbf{y}_{t_{1:j}}\big)$
but an approximation of ${\mathbf E}  \big( (\mathbf{X}_{t_{j}}-
\mathbf{X}_{t_{j-1}})^{2} \big| \mathbf{y}_{t_{1:j}}\big)$ \big(both are
different because ${\mathbf E} ( \mathbf{X}_{t_{j}}-\mathbf{X}_{t_{j-1}} |
\mathbf{y}_{t_{1:j}}) \neq 0$\big). As a consequence of $\,{\mathbf E} \big[{\mathbf E}
\big( (\mathbf{X}_{t_{j}}- \mathbf{X}_{t_{j-1}})^{2} \big|
\mathbf{Y}_{t_{1:j}}\big)\big] = {\mathbf E}  \big(\mathbf{X}_{t_{j}}-
\mathbf{X}_{t_{j-1}}\big)^{2} = \text{Var} \big(\mathbf{X}_{t_{j}}-
\mathbf{X}_{t_{j-1}}\big)$, $\;\hat{\Sigma}_{t_{j}}$ is a descent estimator of
$\Sigma_{t_{j}}$.

\bigskip

\noindent 2) \underline{Time-constant covariance matrices:} If $\Sigma_{t_j}$
is time-constant the first idea is to apply the algorithm (\ref{seqEM1g}) with
the ``constant parameter setting'' $\lambda_{j}= 1 / (j-1)\,$. This corresponds to the global average in (\ref{EM:Estep}) where all observations have equal weights. However, the
situation is different from the classical case in that the ``old'' estimate
$\hat{\Sigma}_{t_{j-1}}$ has in addition some bias due to the use of particles
generated with an estimated covariance instead of the true one. Therefore we
need to put less weight on the first term in (\ref{seqEM1g}). The situation
has been carefully investigated for a similar algorithm in the {i.i.d.}-case
by Capp{\'{e}} and Moulines (2009). Following their recommendation we use in
the time-constant case the on-line algorithm
\begin{equation} \label{seqEM1h}
\hat{\Sigma}_{t_{j}} =
\{1-(j-1)^{-\gamma}\} \, \hat{\Sigma}_{t_{j-1}} + (j-1)^{-\gamma} \, \breve{\Sigma}_{t_{j}}(\omega_{t_j})
\end{equation}
with $\gamma \in (\frac {1} {2},1)$. Capp{\'{e}} and Moulines prove
consistency and asymptotic normality of their estimate for weights
$\lambda_{j}:= \lambda_0 j^{\,-\gamma}$ and $\gamma \in (\frac {1} {2},1)$ and
also for $\gamma=1$ under some restrictions on $\lambda_0$ (Theorem~2).
Furthermore, in their simulations it turned out that a value of $\gamma=0.6$
and $\lambda_0=1$ has lead to good estimates. From our experience we prefer the
choice $\gamma=0.9$ and $\lambda_0=1$ (see Figure~\ref{fig:SimulatedDataBoxPlots_vola}).
Even-Dar and Mansour (2003) obtained an optimal value of about $0.85$ in a
related estimation problem.

\bigskip

\noindent 3) \underline{Time-varying covariance matrices:} If $\Sigma_{t_j}$ is time-varying we use the algorithm (\ref{seqEM1g}) with time-constant
$\lambda_{j}\equiv\lambda$ instead of a decaying $\lambda_j$. The choice of $\lambda$ depends on the smoothness of the true volatility curve. To adapt locally to this smoothness one may either choose a
time varying $\lambda_j$ anyhow (in some way dependent on the data) or use the SAGES procedure (see
Section~\ref{ch:msn:impl} below) where the algorithm is run simultaneously for $L$ different values
of $\lambda$ and the optimal estimate is determined in  each step as a convex combination of these
estimates.

\subsection{Combining the Particle Filter and the Sequential EM-Type Algorithm} \label{ch:msn:ssmEMalg:sum}

To summarize the estimation method for the transaction time volatility and the filtering distribution of the efficient price consists of 3 components:

\begin{figure}[t]
\centering
\includegraphics[width=380pt,keepaspectratio]{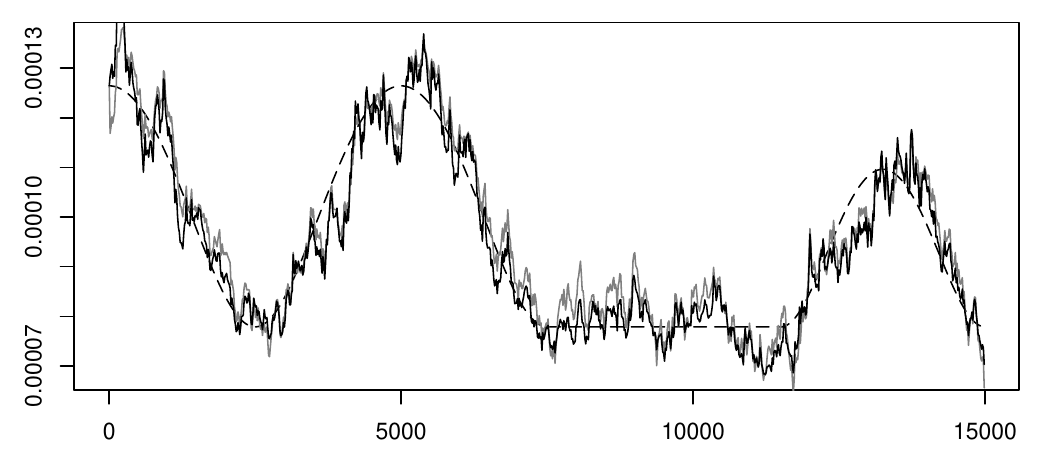}
\includegraphics[width=380pt,keepaspectratio]{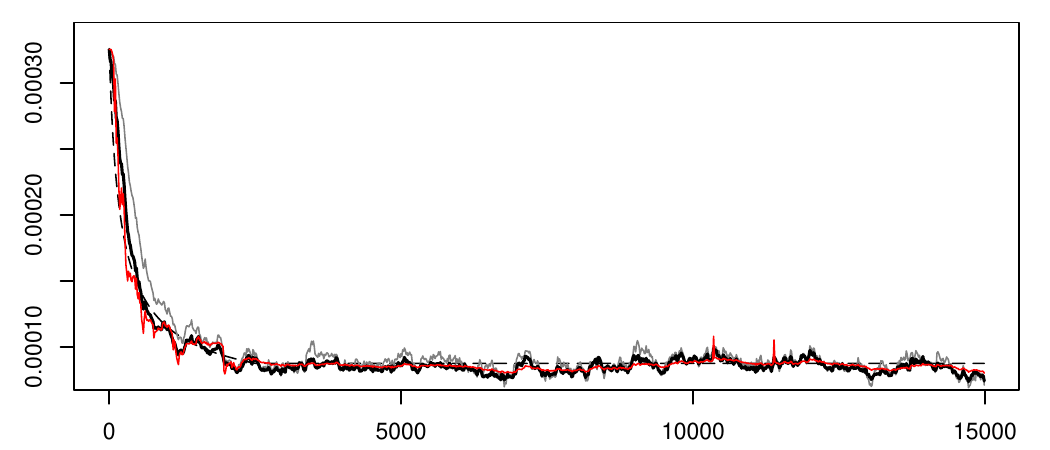}
\caption{\footnotesize Estimation of two time-varying volatility curves given by
the dashed lines based on simulated data. The plots show the estimator $\hat{\Sigma}_{t_j}$ (black line) and a benchmark estimator (light gray line). The second plot also shows an adaptive version of the first estimator based on SAGES $\hat{\Sigma}_{t_j}^{S}$ (red (dark gray) line) described below. For details see
Section~\ref{ch:msn:sim:simdata}. In the first plot this estimator is omitted since it did not lead to additional improvements.}
\label{fig:SimulatedData_allEstimators}
\end{figure}

\begin{enumerate}
  \item[(i)] The state-space model with a new market microstructure noise model and the
      transaction time model for the efficient log-price ((\ref{roundssm:synchObsEquation}) and
      (\ref{roundssm:synchStateEquation}));
  \item[(ii)] A particle filter which sequentially approximates the filtering
  distributions of the efficient log-prices given the observed transaction
  prices (Section~\ref{ch:msn:ssmEMalg:pf});
  \item[(iii)] The on-line EM-type estimator $\hat{\Sigma}_{t_{j}}$ given by (\ref{seqEM1g}) or (\ref{seqEM1h})
  which estimates $\Sigma_{t_j}$ based on the particle approximation of the
  filtering distribution (Section~\ref{ch:msn:ssmEMalg:em}).
\end{enumerate}

A key aspect of the method is the back and forth between the particle filter and the EM-type
estimator. To propagate the particles from time $t_{j}$ to time $t_{j+1}$ the particle filter
requires an estimator of $\Sigma_{t_{j+1}}$ denoted by
$\hat{\Sigma}_{t_{j+1}}^{\text{pf}}$. A simple solution is to use
$\hat{\Sigma}_{t_{j+1}}^{\text{pf}}:=\hat{\Sigma}_{t_{j}}$ from the previous EM-type step. The
EM-type estimator then in turn updates the covariance estimate based on the new particles for time
$t_{j+1}$ generated by the particle filter.

Estimation results of our estimator $\hat{\Sigma}_{t_{j}}$ and a benchmark estimator
(see Section~\ref{ch:msn:sim:simdata}) are presented in
Figure~\ref{fig:SimulatedData_allEstimators}. Details and a discussion
are given in Section~\ref{ch:msn:sim:simdata}.

\section{A Decomposition of Clock Time Volatility\\(from transaction time to clock time)} \label{ch:msn:clock}

%\textbf{\Large From Transaction Time to Clock Time}

\noindent \textbf{The Basic Relationship}

\medskip

\noindent We define the spot volatility in clock time by
\begin{equation*} \label{}
\Sigma^{c}(t) := \lim_{\Delta t \rightarrow 0} \frac
{\text{Var} \big( \mathbf{X}(t+\Delta t) - \mathbf{X}(t)\big)} {\Delta t}\,.
\end{equation*}
For example in the model $d \mathbf{X}(t) \!= \!\tilde{\Gamma}(t) \, d \mathbf{W}_{t}$ with a Brownian motion $\mathbf{W}_{t}$ (multivariate with independent components) we have $\text{Var} \big( \mathbf{X}(t+\Delta t) - \mathbf{X}(t)\big) = \int_{t}^{t+\Delta t} \tilde{\Gamma}  (s) \tilde{\Gamma}  (s)' \,ds$ and therefore $\Sigma^{c}(t) = \tilde{\Gamma}  (t) \tilde{\Gamma} (t)'$. At the end of this section we indicate how estimation can be performed directly in this model with a particle filter.

In this paper we merely advocate the model $\mathbf{X}_{t_j} = \mathbf{X}_{t_{j-1}} + \Gamma (t_{j}) Z_j$ with $Z_j \stackrel{iid}{\sim} \mathcal{N}(0,I)$ (see (\ref{roundssm:synchStateEquation})) which can be written in the form $d \mathbf{X}(t) \!= \!\Gamma(t) \, d \mathbf{W}_{N(t)}$ with $N(t)= \sum_j I_{[t_j,\infty)}(t)$. If we now assume that the observation times $t_j$ are realizations of a stochastic point process with intensity function $\lambda_{I}(t)$ (transaction rate) and $N(\cdot)$ is independent of $\mathbf{W} (\cdot)$ (i.e. the dependence only enters via $\Gamma (t_{j})$) then
\begin{equation*} \label{}
\text{Var} \big( \mathbf{X}(t+\Delta t) - \mathbf{X}(t)\big) = \int_{t}^{t+\Delta t}\! \!\Sigma (s) \, \lambda_{I}(s) \, ds \quad \mbox{with}\quad \Sigma (t) := \Gamma (t) \Gamma (t)'
\end{equation*}
and therefore
\begin{equation} \label{RelationVolatilities}
\Sigma^{c}(t) \!=\! \Sigma(t)\, \lambda_{I}(t).
\end{equation}
Heuristically this reads as ``variance per time unit = variance per transaction $\times$ expected number of transactions per time unit''. This is a decomposition of continuous time volatility which provides a deeper understanding of volatility. An example is given below. We now use this relation for the estimation of $\Sigma^{c}(t)$.

We mention that both curves can be identified if in addition to the $X_{t_{j}}$ also the random times $t_j$ are observed (which is fulfilled in our setting). A proof of consistency of estimates of $\lambda_I(t)$ (e.g. of $\hat{\lambda}_{I}(t_j)$ from below) would require an in-fill asymptotic setting where $N(t)$ has the intensity function $\lambda_I \big(\frac {t} {T}\big)$.

Models with random time changes are common in finance (cf. Clark 1973; An{\'{e}} and Geman 2000; Plerou et al. 2001; Howison and Lamper 2001; Gabaix et al. 2003). The process of random times (here the $t_j$) is often called directing process and the process $X (t)$ is called subordinated to the directing process. Another example is where $N(t)$ is replaced by the accumulated traded volume.

\bigskip
\medskip

\noindent \textbf{A New Estimator for the Clock Time Spot Volatility}

\medskip

We now use the relation (\ref{RelationVolatilities}) for the estimation of $\Sigma^{c}(t)$.  An obvious estimate of the intensity would be  $\hat{\hat{\lambda}}_{I}(t_j) : =   |\{\ell\,:\,t_j - \Delta t < t_{\ell} \le t_j\}| \, / \, \Delta t$ with some $\Delta t$.

Here we advocate a different estimation method of the intensity function
$\lambda_{I}(t)$ which is closer related to our on-line scheme, namely the
estimation of $\lambda_{I}(t)$ by the inverse of the averaged duration times, leading to the alternative estimate
\begin{equation} \label{AlternativeClockEstimate}
\hat{\Sigma}_{\text{alt}}^{c}(t_j) : =\! \hat{\Sigma}_{t_{j}}\, \hat{\lambda}_{I}(t_j) \quad \mbox{with}\quad \hat{\lambda}_{I}(t_j) := \frac {1} {\bar{\delta}_{j}} \, \textsf{c}_{bc}
\end{equation}
and $\hat{\Sigma}_{t_{j}}$ as in (\ref{seqEM1g}) where $\bar{\delta}_{j}$ is defined by the recursion
\begin{equation} \label{seq-durations}
\bar{\delta}_{j} = (1-\lambda_j) \, \bar{\delta}_{j-1} +
\lambda_j \, \big(t_{j} - t_{j-1} \big) \quad \mbox{with} \quad \bar{\delta}_{2} =t_{2} - t_{1}.
\end{equation}
(the notation $\hat{\Sigma}_{\text{alt}}^{c}$ means ``alternative'' estimate in comparison to the more classical estimate defined below). $\textsf{c}_{bc}$ is a bias correction due to the fact that $\mean \frac {1} {X} \! \neq \! \frac {1} {\mean X}$. A second order Taylor-expansion leads to $\mean \frac {1} {X} \approx \frac {1} {\mean X} \big(1 + \frac {\var(X)} {(\mean X)^2}\big)$ and we therefore use the above estimate with $\textsf{c}_{bc} = (1 + \hat{d})^{-1}$ where $\hat{d}$ is an estimate of $\frac {\var(\bar{\delta})} {(\mean \bar{\delta})^2}$.

We mention that the intensity $\lambda_{I}(t)$ of the point
process often changes considerably over time thus requiring a large
value of $\lambda$ while $\Sigma (t)$ usually is more smooth. For that reason we use different step sizes $\lambda$ for
the estimators $\hat{\Sigma}_{t_j}$ and $\bar{\delta}_{j}$ (cf.
Section~\ref{ch:msn:sim:realdata}).

\bigskip
\medskip

\noindent \textbf{Estimation of the Clock Time Spot Volatility without Time Change}

\medskip

We now define the estimator of the clock time volatility in the classical model $d \mathbf{X}(t) \!= \!\tilde{\Gamma}(t) \, d \mathbf{W}_{t}$ with the microstructure noise model from above. If we replace $\mathbf{X}_{t_j}$ by $\mathbf{X}(t_j)$ we obtain
almost the same state space model as in (\ref{roundssm:synchObsEquation}) and
(\ref{roundssm:synchStateEquation}) but with a modified variance of the
transition distribution which is now given by
\begin{equation}\label{retdistr:clocktime}
p\big(\mathbf{x}_{t_j}\big| \mathbf{x}_{t_{j-1}} \big) =
\mathcal{N}\big(\mathbf{x}_{t_j} \big| \mathbf{x}_{t_{j-1}};
|t_{j}-t_{j-1}| \; \Sigma^{c} (t_j) \big).
\end{equation}
This is the only change needed in the state-space model (\ref{roundssm:synchObsEquation}),
(\ref{roundssm:synchStateEquation}). As an estimate $\hat{\Sigma}_{t_j}^{c}$ we can use the on-line estimates
(\ref{seqEM1h}) and (\ref{seqEM1g}) but now with the

update matrix $\breve{\Sigma}_{t_j}(\omega_{t_j})$ replaced by
\begin{equation} \label{SigmaBreveC}
\breve{\Sigma}_{t_j}^{c}(\omega_{t_j}^c) := \sum_{i=1}^N \omega_{t_j}^{ci}
\frac{ \big(\mathbf{x}_{t_j}^{ci} - \mathbf{x}_{t_{j-1}}^{ci}\big) \big(\mathbf{x}_{t_j}^{ci} -
\mathbf{x}_{t_{j-1}}^{ci}\big)'} {|t_{j}-t_{j-1}|}
\end{equation}
based on the modified filtering particles $\{\mathbf{x}_{t_{j-1:j}}^{ci},  \omega_{t_{j}}^{ci}
\}_{i=1}^N$.

\bigskip

\begin{figure}[t]
\centering
\includegraphics[width=0.77\textwidth,keepaspectratio]{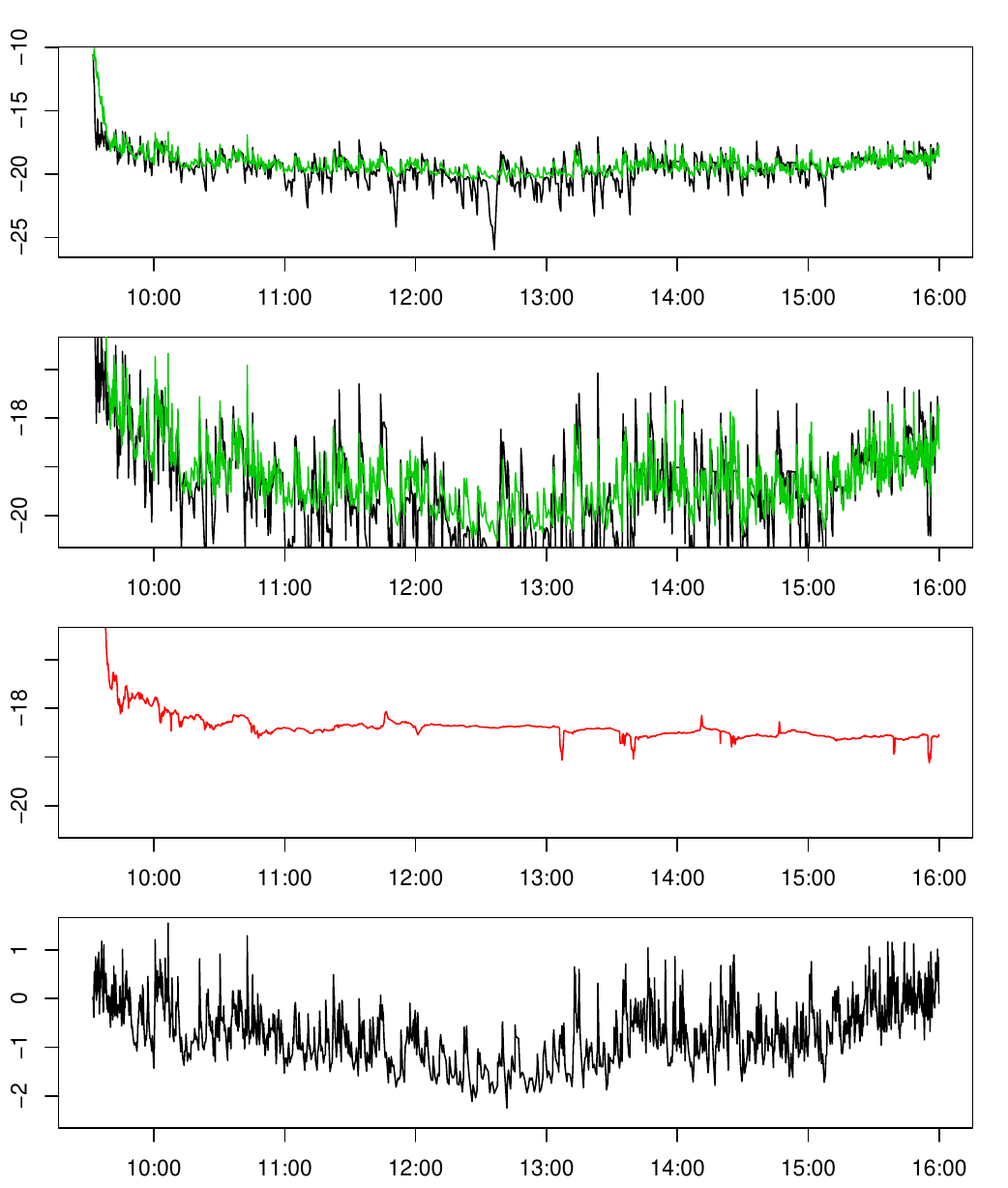}
\caption{\footnotesize
Real data example: Estimation of time-varying spot volatility in clock time
based on the transactions of symbol C for the 3rd September 2007.
The first and second plot give the log volatility estimators $\hat{\Sigma}^{cS}_{t_j}$
(black line) and $\hat{\Sigma}_{\text{alt}}^{cS}(t_j)$ (green (gray) line) where the second plot uses the same scaling as the following plots; the green (gray) estimator is the sum of the log volatility estimator in transaction time $\hat{\Sigma}^{S}_{t_j}$ (third plot) and the log trading intensity $\hat{\lambda}_{I}(t_j)$ (last plot). The superscript `S' denotes the SAGES - version - see Section~\ref{ch:msn:impl}.}
\label{fig:RealDataClock_allEstimators}
\end{figure}

We conclude this section with a heuristics on the relation between the two estimates: Suppose the same stepsize $\lambda$ were used for the calculation of $\hat{\Sigma}_{t_j}$ and $\bar{\delta}_{j}$. We then had with (\ref{seqEM1gClosedForm})
\begin{align*} \label{}
\hat{\Sigma}_{\text{alt}}^{c}(t_j) & = \frac {\hat{\Sigma}_{t_j}}
{\bar{\delta}_{j}}  \, \textsf{c}_{bc} = \frac {\lambda \, \sum_{k=0}^{j-3}
(1-\lambda)^{k} \;
\breve{\Sigma}_{t_{j-k}}(\omega_{t_{j-k}}) \, + \,
(1-\lambda)^{j-2} \; \breve{\Sigma}_{t_{2}}(\omega_{t_2})}
{\lambda \, \sum_{k=0}^{j-3}
(1-\lambda)^{k} \;  \big(t_{j-k} - t_{j-k-1} \big) + (1-\lambda)^{j-2} \;  \big(t_{2} - t_{1} \big)}  \, \textsf{c}_{bc}.
\end{align*}
Since $\breve{\Sigma}_{t_{\ell}}(\omega_{t_{\ell}}) \approx
\big(t_{\ell} - t_{\ell-1} \big) \,
\breve{\Sigma}_{t_{\ell}}^{c}(\omega_{t_{\ell}}^c)$ the estimator is of the
form
\begin{equation*} \label{}
\hat{\Sigma}_{\text{alt}}^{c}(t_j) \approx \frac {\sum_{k=0}^{j-2} w_{k}
\breve{\Sigma}_{t_{j-k}}^{c}(\omega_{t_{j-k}}^c)} {\sum_{k=0}^{j-2} w_{k}} \, \textsf{c}_{bc} \,,
\end{equation*}
that is $\hat{\Sigma}_{\text{alt}}^{c}(t_j)$ is a weighted average of the
$\breve{\Sigma}_{t_{\ell}}^{c}(\omega_{t_{\ell}}^c)$ and therefore a similar estimator as in the
clock time model. The ``$\approx$'' signs stem from the fact that in $\breve{\Sigma}_{t_{\ell}}(\omega_{t_{\ell}})$ and $\breve{\Sigma}_{t_{\ell}}^{c}(\omega_{t_{\ell}}^c)$ different particle filters for different models are used. This effect usually cannot be neglected.

\bigskip
\newpage

\noindent \textbf{Decomposing Clock Time Volatility}
\medskip

Figure~\ref{fig:RealDataClock_allEstimators} shows an example based on real data which is discussed in detail in Section~\ref{ch:msn:sim:realdata}. We have used a log plot in order to demonstrate the influence of the two curves in the decomposition (\ref{AlternativeClockEstimate}) which now becomes
\begin{equation*} \label{}
\log \hat{\Sigma}_{\text{alt}}^{cS}(t_j) = \log \hat{\Sigma}^{S}_{t_{j}} + \log \hat{\lambda}_{I}(t_j)
\end{equation*}
where the superscript `S' denotes the adaptive SAGES - version of the estimators described below. The first and second plot give the log of the volatility estimators $\hat{\Sigma}^{cS}_{t_j}$
(black line) and $\hat{\Sigma}_{\text{alt}}^{cS}(t_j)$ (green (gray) line) where the second plot uses the same scaling as the following plots; the green (gray) estimator is the sum of the log  volatility estimator in transaction time $\hat{\Sigma}^{S}_{t_j}$ (third plot) and the log trading intensity $\hat{\lambda}_{I}(t_j)$ (last plot).

The decomposition of clock time volatility into transaction time volatility and trading intensity (i.e. the additive decomposition of the green (gray) curve of Figure~\ref{fig:RealDataClock_allEstimators} into the two lower plots) reveals that the typical fluctuation of clock time volatility is mainly due to the fluctuation of the trading intensity while the transaction time volatility in this example is almost constant after 11:00. The typical U-shape of clock time volatility is visible - but it is more a pattern of the trading intensity and less of the transaction time volatility. More precisely, the decrease of volatility between 9:30 and 12:30 is a feature of both curves while the increase of volatility between 12:30 and 16:00 is only a feature of trading intensity.

It is worth mentioning that the black and the green (gray) estimators in Figure~\ref{fig:RealDataClock_allEstimators} coincide in
magnitude (which was not clear beforehand since different models and different particle filters are used). Figure~\ref{fig:RealDataClock_allEstimatorsA} below compares the estimates for a small time period.

\section{Modifications, Adaptation, and Implementation} \label{ch:msn:impl}

\subsection*{Using returns of lag $k$}

There exist some objections against the use of ultra-high frequency data at the finest level available. Here we show how the method can be used on a coarser scale, together with a few comments on the situation.

It is common to use returns of lag $k$ instead of lag $1$, the main reason being that microstructure noise is smaller in the averaged data. Our efforts in this paper were to construct a better microstructure noise model and to remove the microstructure noise by a particle filter prior to the calculation of the volatility estimate. This allows us to investigate ultra-high frequency data at a finer level. An application is the pricing of high frequency options which can be traded until a few seconds to maturity.

On the other hand it is likely that our microstructure noise model still is not perfectly specified. For that reason one may still want to use returns of lag $k$ instead of lag $1$. In our setting this can be accomplished by using
\begin{equation*} \label{}
\breve{\Sigma}_{t_{j}}(\omega_{t_j}) := \frac {1} {k}\sum_{i=1}^N \omega_{t_{j}}^i
\big(\mathbf{x}_{t_{j}}^i-\mathbf{x}_{t_{j-k}}^i\big)
\big(\mathbf{x}_{t_{j}}^i-\mathbf{x}_{t_{j-k}}^i\big)',
\quad
\breve{\Sigma}_{t_j}^{c}(\omega_{t_j}^c) := \sum_{i=1}^N \omega_{t_j}^{ci}
\frac{ \big(\mathbf{x}_{t_j}^{ci} - \mathbf{x}_{t_{j-k}}^{ci}\big) \big(\mathbf{x}_{t_j}^{ci} -
\mathbf{x}_{t_{j-k}}^{ci}\big)'} {|t_{j}-t_{j-k}|}
\end{equation*}
in (\ref{EstimateSigma4}) and (\ref{SigmaBreveC}) (with the recursions (\ref{seqEM1h}) and (\ref{seqEM1g}) as before). We think that this is in particular important for the continuous time estimator $\breve{\Sigma}_{t_j}^{c}(\omega_{t_j}^c)$ which explodes for very small values of $|t_{j}-t_{j-k}|$ - this happens more often for $k=1$. As a consequence of the larger lag also a larger stepsize is necessary - also in the SAGES procedure introduced below.

Note, that the estimate is still updated with each new observation. Furthermore, the conditional distribution of the state is calculated with a new observation. When implementing the above estimate, special care is needed if a resampling step is carried out between $t_{j}$ and $t_{j-k}$.

In a correctly specified model (where in particular microstructure noise is specified correctly) the variance and the mean squared error get smallest for lag 1. On the other hand the bias due to a misspecified microstructure noise model gets smaller with larger lag. In principle one may test the quality of the microstructure noise model by comparing the level of the estimates for different lags. However, this topic is beyond the scope of this paper.

\subsection*{Improving estimates in the presence of diurnal patterns}

Diurnal patterns like the strong decrease of the volatility at the beginning of the day in Figure~\ref{fig:RealData_allEstimatorsStartValue} create problems in that an unadjusted look-back local estimator overestimates the target. Similarly, when the volatility is rising, a look-back local estimator underestimates the target. The SAGES procedure below reduces the effect but the problem in principle stays the same. In the present setting the situation is even more critical since those poor volatility estimates are used afterwards in the particle filter.

A common advice is to use a batch of days to estimate the mean diurnal volatility pattern over small blocks of time, scale out the pattern yielding diurnally-adjusted data, estimate the local object of interest on the adjusted data, and then rescale back to account for the diurnal pattern.
Such a modification can also be applied with the procedure of this paper.

The key difference to other situations is that our volatility estimator consists of the product of two curves corresponding to the decomposition (\ref{RelationVolatilities}). Both curves can be identified and both curves can be adjusted for diurnal patterns. As an example we argue in Section~\ref{ch:msn:sim:realdata} that (at least for the data set analyzed there) the well known
U-shape effect at lunchtime is a diurnal pattern merely of the trading intensity $\lambda_{I}(t)$ and not of the trading time volatility $\Sigma(t)$.

When rescaling the estimate of $\Sigma(t)$ special care is needed in order not to affect the microstructure noise model: Suppose the mean diurnal
volatility pattern of the s-th component is $\sigma_{0s}^{2} (t)$. Let $V_0 (t) := {\rm diag} \{\sigma_{01} (t),\ldots,\sigma_{0S} (t)\}$. Instead of rescaling the observations we use the rescaled (unobserved) state-variable $\tilde{X}_{t_j} : = V_0 (t_j)^{-1} X_{t_j}$. Provided that the difference between $\sigma_{0s} (t_j)$ and $\sigma_{0s} (t_{j-1})$ is negligible we then can use instead of (\ref{roundssm:synchObsEquation}) and (\ref{roundssm:synchStateEquation}) the modified state space model
\begin{eqnarray}
\label{roundssm:synchObsEquation1}
p\big(\mathbf{y}_{t_j}\big| \mathbf{y}_{t_{1:j-1}}, \exp(\tilde{\mathbf{x}}_{t_j})\big)
& \propto & \mathbf{1}_{\mathbf{A}_{t_j}}\big(\exp(V_0 (t_j) \tilde{\mathbf{x}}_{t_j})\big) \, \mathbf{1}_{\mathcal{Y}} (\mathbf{y}_{t_j})\\
& = & \mathbf{1}_{V_0 (t_j)^{-1} \log \mathbf{A}_{t_j}}\big(\tilde{\mathbf{x}}_{t_j}\big)  \, \mathbf{1}_{\mathcal{Y}} (\mathbf{y}_{t_j})\quad \mbox{a.s.},\\
\label{roundssm:synchStateEquation1}
p\big(\tilde{\mathbf{x}}_{t_j}\big| \tilde{\mathbf{x}}_{t_{j-1}} \big) &=&
\mathcal{N}\big(\tilde{\mathbf{x}}_{t_j} \big| \tilde{\mathbf{x}}_{t_{j-1}};
V_0 (t_j)^{-1} \Sigma_{t_{j}} V_0 (t_j)^{-1}\big)
\end{eqnarray}
for estimation. This means we can run the whole procedure in exactly the same way where $\log \mathbf{A}_{t_{j}}$ in the proposal distribution and in the importance weights (Proposition 1) is replaced by $V_0 (t_j)^{-1} \log \mathbf{A}_{t_{j}}$. The resulting volatility estimator $\hat{\Sigma}_{t_{j}}$ then is an estimator of $V_0 (t_j)^{-1} \Sigma_{t_{j}} V_0 (t_j)^{-1}$, i.e. we finally use $V_0 (t_j)\hat{\Sigma}_{t_{j}} V_0 (t_j)$ as an estimator of $\Sigma_{t_{j}}$.

Rescaling the estimator of $\lambda_{I}(t)$ in (\ref{RelationVolatilities}) is much simpler: Suppose the mean diurnal intensity pattern is $\lambda_{0}(t)$. A natural recursive estimator then is
\begin{equation*} \label{}
\hat{\lambda}_{I}(t_j) := \lambda_{0}(t_j) \, \frac {1} {\bar{\delta}_{j}} \, \textsf{c}_{bc}
\end{equation*}
where $\bar{\delta}_{j}$ is defined by the recursion
\begin{equation*} \label{}
\bar{\delta}_{j} = (1-\lambda_j) \, \bar{\delta}_{j-1} +
\lambda_j \,  \big(t_{j} - t_{j-1}\big) \, \lambda_{0}(t_j)
\end{equation*}
and $\textsf{c}_{bc}$ is the corresponding bias correction.

\subsection*{Step size selection}

In the time-constant case we use the decreasing step size $\lambda_j = (j-1)^{-0.9}$ as proposed in
Section~\ref{ch:msn:ssmEMalg:em}. This choice is empirically justified (see
Figure~\ref{fig:SimulatedDataBoxPlots_vola}).

The step size in the time-varying case is data dependent and can be obtained through the following
procedure: The mean squared error of $\hat{\Sigma}_{t_j}$ is minimized with respect to
$\lambda$ by the cross-validation type criterion
\begin{equation} \label{CrossValidation}
    {\rm crit} (\lambda) := \sum_{j=2}^{T-1} \big(\hat{\Sigma}_{t_j} -
    \breve{\Sigma}_{t_{j+1}}(\omega_{t_{j+1}})\big)^2.
\end{equation}
This cannot be done on-line. In practice, one will use in an on-line
setting a $\lambda$ from past experience with similar data sets. The
expectation of the above criterion is approximately
\begin{equation*} \label{}
    \sum_{j=2}^{T-1} \Big[\big({\mathbf E} \hat{\Sigma}_{t_j} -
    \Sigma_{t_j}\big)^{2} + {\text{Var}} \big(\hat{\Sigma}_{t_j}\big) +
    {\text{Var}} \big(\breve{\Sigma}_{t_{j+1}}(\omega_{t_{j+1}})\big)\Big].
\end{equation*}
Because the last term does not depend on $\lambda$ we correctly
minimize the approximate mean squared error.

\subsection*{Adaptive step size selection using SAGES}

To adaptively select non-constant step sizes $\lambda_j$ in the time-varying case we propose to use
spatially aggregated exponential smoothing (SAGES) developed by Chen and Spokoiny (2009). In our
setting the SAGES method works as follows. The basic idea is to run $L$ volatility estimators
$\hat{\Sigma}_{t_{j}}^{\ell}$ in parallel with different step sizes $\lambda^1 > \lambda^2 > \ldots >
\lambda^L$. The resulting SAGES estimate $\hat{\Sigma}_{t_{j}}^{S}$ is then a convex combination of
these estimators. In practice we have, say, $L=15$ which implies that the computational offset is
minimal. In fact, only the recursion (\ref{seqEM1g}) needs to be computed $L$ times with different
step sizes.

For every time step $j$ the SAGES estimate $\hat{\Sigma}_{t_{j}}^{S}$ is obtained from the
estimators $\hat{\Sigma}_{t_{j}}^{\ell}$, $\ell=1,\ldots,L$, through the following recursion.
\begin{itemize}
\item[(i)] Set $\hat{\Sigma}_{t_{j}}^{S,1} = \hat{\Sigma}_{t_{j}}^{1}$
\item[(ii)] {For $\ell=2,\ldots,L$}: \; Compute
\begin{equation*}
\hat{\Sigma}_{t_{j}}^{S,\ell} = \left( \frac{\gamma_\ell}{\hat{\Sigma}_{t_{j}}^{\ell}} +
\frac{1-\gamma_\ell}{\hat{\Sigma}_{t_{j}}^{S,\ell-1}} \right)^{\!-1},
\end{equation*}
where
\begin{equation*}
\gamma_\ell = K\left( \frac{1}{\kappa_{\ell-1} \lambda^\ell} \,\mathcal{K}( \hat{\Sigma}_{t_{j}}^{\ell},
\hat{\Sigma}_{t_{j}}^{S,\ell-1}) \right)
\end{equation*}
with kernels $K(u) = \{1-(u-1/6)^{+}\}^{+}$ and $\,\mathcal{K}(\Sigma, \tilde{\Sigma}) = -0.5
\big\{\log(\Sigma / \tilde{\Sigma}) + 1 - \Sigma / \tilde{\Sigma} \big\}$.
\item[(iii)] Obtain the SAGES estimate $\hat{\Sigma}_{t_{j}}^{S} = \hat{\Sigma}_{t_{j}}^{S,L}$.
\end{itemize}
Note that this method can be applied completely on-line. The parameters $\kappa_{1}, \kappa_{2},
\ldots, \kappa_{\ell-1}$ are critical values (independent of the time step $j$) which can be
calculated beforehand through a Monte Carlo simulation. Note that SAGES is a univariate method. For
a more detailed description and a theoretical analysis of the SAGES method see Chen and Spokoiny
(2009).

\subsection*{Implementing the algorithm}

The particle filter uses the following steps for $j=2,\ldots,T$ (see Proposition~\ref{msn:prop1})
\begin{itemize}
\baselineskip1.1em
\itemsep-0.03cm
\item {For $i=1,\ldots,N$}: $\;$
                \begin{itemize}
                \baselineskip1.1em
                \itemsep-0.03cm
                        \item Generate $\mathbf{x}_{t_j}^i$ from the optimal proposal
                            $\mathcal{N}(\mathbf{x}_{t_{j}}| \mathbf{x}_{t_{j-1}}^{i};
                            \hat{\Sigma}_{t_{j}}^{\text{pf}})\big|_{ \log
                            \mathbf{A}_{t_{j}}}$ with $\hat{\Sigma}_{t_{j}}^{\text{pf}} =
                            \hat{\Sigma}_{t_{j-1}}$.
                        \item Compute the importance weight $\breve{\omega}_{t_{j}}^i$ as
                            in (\ref{IScomputationEq}). If $S=1$ this is given by
                        \begin{equation*} \label{}
                        \breve{\omega}_{t_{j}}^i \propto
                        \omega_{t_{j-1}}^i \Big\{\Phi\big(\sup \log A_{t_{j}}|
                        x^i_{t_{j-1}}; \hat{\Sigma}_{t_{j}}^{\text{pf}}\big)
                        -\Phi\big(\inf \log A_{t_{j}}| x^i_{t_{j-1}};
                        \hat{\Sigma}_{t_{j}}^{\text{pf}}\big) \Big\}.
                        \end{equation*}
                \end{itemize}
                \item {For $i=1,\ldots,N$}: $\;$  Normalize the importance weight
                    $\omega_{t_j}^i = \breve{\omega}^i_{t_j} / \sum_{k=1}^N
                    \breve{\omega}^k_{t_j}$.
        \item If the effective sample size $\text{ESS}(\{\omega_{t_j}^i \}_{i=1}^N) < c N$
            (with say $c=0.2$), then resample the particles using, for instance, the residual
            resampling scheme (Douc et al. 2005).
        \item Update the estimator $\hat{\Sigma}_{t_{j-1}}$ according to (\ref{seqEM1h}) or
            (\ref{seqEM1g}).
\end{itemize}

Overall the algorithm is easy to implement in a few lines. It is computationally
efficient because the complexity of one iteration is linear in the number of particles $N$. In
addition resampling is required only rarely because the optimal proposal is used. In our
applications resampling was carried out only about every 15th iteration using a threshold for the
effective sample size of $c=0.2$. As a result of the efficiency of our particle filter, the number
of particles $N$ is not a critical quantity. Typically, about 500 particles suffice to achieve a
sufficient precision (see Figure \ref{fig:SimulatedDataBoxPlots_vola}).

Note that in the multivariate case the sampling from the optimal proposal and the evaluation of the
importance weights is nontrivial. However, both the sampling from and the evaluation of a truncated
normal distribution are standard problems in statistics which have been discussed extensively in
the literature. Relevant references for the sampling problem are Geweke (1991) and Robert (1995).
More recent approaches based on Gibbs sampling are described by Kotecha and Djuric (1999) and
Rodriguez-Yam et al. (2004). Also for the numerical approximation of multivariate (rectangular)
normal probabilities several efficient methods have been proposed for instance by Genz (1992, 2004)
and Joe (1995).

\subsection*{Initialization}

Our experience from many data sets is that the algorithm stabilizes quickly provided that
reasonable starting values are used -- e.g. $\hat{\Sigma}_{t_2}$ may be chosen as yesterday's starting volatility or yesterday's ending volatility, after adjustment for the magnitude of the overnight close-to-open jump. The particle filter is started by simulating the $x_{t_{1},s}^i$
such that the $\exp(x_{t_{1},s}^i)$ are uniformly distributed on $A_{t_1,s}$. In order to exclude
the effect of starting values we have used in the simulations (except from
Figure~\ref{fig:SimulatedDataBoxPlots_vola}) the true matrix $\Sigma_{t_2}$ as the starting value
(i.e. $\hat{\Sigma}_{t_{2}}^{\text{pf}} = \hat{\Sigma}_{t_2} =\Sigma_{t_2}$).

\section{Simulations and Applications} \label{ch:msn:sim}

\subsection{Results for Simulated Data} \label{ch:msn:sim:simdata}

\begin{figure}
\centering
\includegraphics[width=430pt,keepaspectratio]{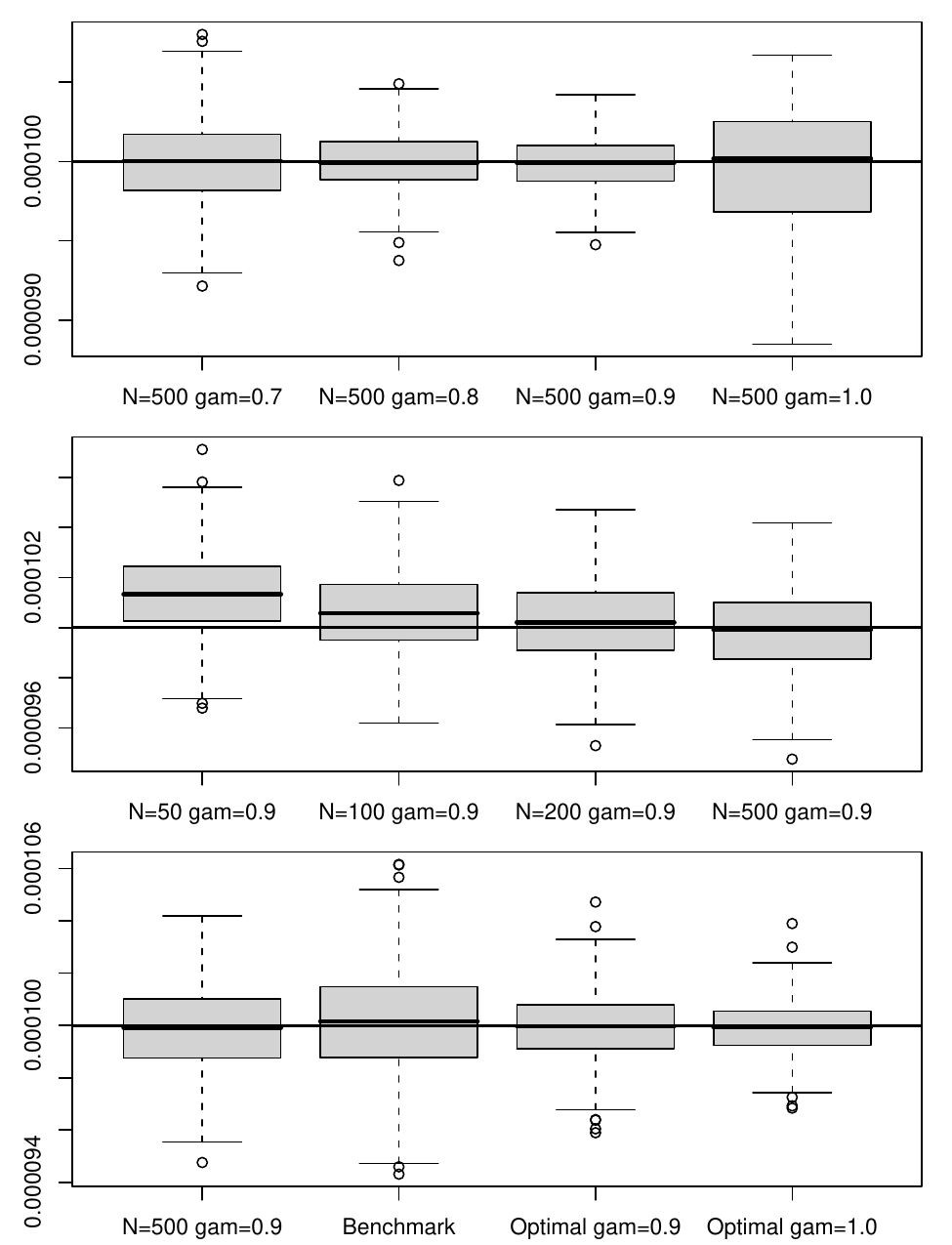}
\caption{\footnotesize
Box plots for the estimation of a time-constant volatility
based on simulated data (5,000 transactions). The estimator
(\ref{seqEM1h}) is applied with different numbers of particles $N$ and
different $\gamma$ and compared to the  benchmark estimator and the optimal
estimator (not available in practice). The box plots are based on 500
independent runs.}
\label{fig:SimulatedDataBoxPlots_vola}
\end{figure}

\subsubsection*{Estimation of time-constant spot volatility}
We first consider the estimation of time-constant spot volatility. An efficient log-price process
is simulated from $t_1$ to $t_{5000}$ with squared volatility equal to $\Sigma_t = 0.0001^2$. The
initial efficient price $\exp(X_{t_1})$ is sampled from a uniform distribution on $[50-0.005,
50+0.005)$. The transaction prices are obtained by rounding the efficient prices to the nearest
cent (see Example 1 (i) in Section~\ref{ch:msn:msnmodel}).
 The algorithm for time-constant spot
volatility estimation (\ref{seqEM1h}) is applied with different numbers of particles $N$ and
different values of $\gamma$. The initial value $\hat{\Sigma}_{t_{2}}^{\text{pf}} =
\hat{\Sigma}_{t_2}$ is drawn from a uniform distribution on $(0.00006^2, 0.00014^2)$ which is quite
uninformative. For comparison the results of two benchmark algorithms are also reported. The first
benchmark method (``Benchmark'' in Figure \ref{fig:SimulatedDataBoxPlots_vola}) is a recursive
estimator with a simpler microstructure noise correction. It is related to the method in Zumbach et
al. (2002) and it is based on the market microstructure model $\log Y_{t_j} = X_{t_j} + U_{t_j}$,
where the noise variables $U_{t_j}$ are {i.i.d.} with $\text{Var} \ U_{t_j} = \eta^2$. The
recursive estimator is given by
\begin{equation}\label{benchmarkestimator}
\hat{\Sigma}_{t_j}^{\text{B}} := \big\{1-\frac {1} {j-1}\big\}
\big(\hat{\Sigma}_{t_{j-1}}^{\text{B}} + \text{max}\{0, 2 \hat{\eta}_{t_{j-1}}^2\}\big)
+ \frac {1} {j-1} \ (\log y_{t_j}-\log y_{t_{j-1}})^2 - \text{max}\{0, 2 \hat{\eta}_{t_j}^2\}
\end{equation}
where $\hat{\eta}_{t_j}^2 := \{1-\frac {1} {j-2}\} \hat{\eta}_{t_{j-1}}^2 \!-\! \frac {1} {j-2} \
\big(\log y_{t_j}\!-\!\log y_{t_{j-1}}\big) \big(\log y_{t_{j-1}}\!-\!\log y_{t_{j-2}}\big)$ (here
$\frac {1} {j-2}$ is used instead of $\frac {1} {j-1}$ because the algorithm starts one time point
later). The term $\,\text{max}\{0,2 \hat{\eta}_{t_j}^2\}\,$ corrects for the market microstructure
noise. This follows from the fact that
\[
\text{Cov}\big(\log Y_{t_j}-\log Y_{t_{j-1}}, \log Y_{t_{j-1}}-\log Y_{t_{j-2}}\big) = -
\eta^2.
\]
The second benchmark method is, in some sense, the optimal estimator (``Optimal'' in
Figure~\ref{fig:SimulatedDataBoxPlots_vola}). It is unavailable in practice because it uses the
latent efficient log-prices. It is computed analogous to (\ref{seqEM1h}) but instead of the
particles it employs the efficient log-prices leading to
\begin{equation*}\label{optimalestimator}
\hat{\Sigma}^{\text{Opt}}_{t_j} = \{1 - (j-1)^{-\gamma}\}
\hat{\Sigma}^{\text{Opt}}_{t_{j-1}} + (j-1)^{-\gamma} (x_{t_j}-x_{t_{j-1}})^2.
\end{equation*}

\noindent The simulation results are given in terms of box plots which are obtained by 500
independent runs (Figure~\ref{fig:SimulatedDataBoxPlots_vola}). The box plots suggest that our
volatility estimator is asymptotically unbiased and that $\gamma = 0.9$ is a reasonable value. We
can also conclude that about 500 particles are sufficient which makes our algorithm computationally
efficient and suitable for real-time applications. In addition, it can be observed that the
benchmark estimator has a larger variance than our estimator.

\subsubsection*{Estimation of time-varying spot volatility}

We now compare our estimator for time-varying spot volatility $\hat{\Sigma}_{t_{j}}$ defined in
(\ref{seqEM1g}) with a benchmark estimator. The efficient log-prices are generated with respect to
the time-varying volatility given by the gray dashed lines in
Figure~\ref{fig:SimulatedData_allEstimators}. The first case (upper plot) is more challenging while
the second case (lower plot) is more realistic for a volatility curve in transaction time - see the
real data example in Figure~\ref{fig:RealData_allEstimatorsStartValue}. In both cases we use for
the initial price $\exp(X_{t_1}) \sim \mathcal{U}[50-0.005, 50+0.005)$. Again transaction prices
(observations) are obtained by rounding the efficient prices to the nearest cent. 15,000
transactions are generated which is typical for one trading day of a liquid stock. The particle
filter is applied with $N = 500$ particles. Our estimator $\hat{\Sigma}_{t_{j}}$ uses the constant
step size $\lambda$ obtained by minimizing (\ref{CrossValidation}). Analogous to
(\ref{benchmarkestimator}) we consider the benchmark estimator given by
\begin{equation} \label{msn:recursiveBenchTV}
\hat{\Sigma}_{t_{j}}^{\text{B}} := \{1-\lambda\}
\big(\hat{\Sigma}_{t_{j-1}}^{\text{B}} + \text{max}\{0, 2 \hat{\eta}_{t_{j-1}}^2\}\big)
+ \lambda \big(\log y_{t_j} - \log y_{t_{j-1}}\big)^2 -
\text{max}\{0, 2 \hat{\eta}_{t_j}^2\}
\end{equation}
with $\hat{\eta}_{t_j}^2 := \{1-\frac {1} {j-2}\} \hat{\eta}_{t_{j-1}}^2 \!-\! \frac {1} {j-2} \
\big(\log y_{t_j}\!-\!\log y_{t_{j-1}}\big) \big(\log y_{t_{j-1}}\!-\!\log y_{t_{j-2}}\big)$.
$\lambda$  is obtained by minimizing the criterion
\begin{equation} \label{CrossValidationBench}
\sum_{j=2}^{T-1} \big(\hat{\Sigma}_{t_{j}}^{\text{B}} + \text{max}\{0, 2
\hat{\eta}_{t_{j}}^2\} - (\log y_{t_{j+2}} - \log y_{t_{j+1}})^2 \big)^2
\end{equation}
\big(the terms $\hat{\Sigma}_{t_{j}}^{\text{B}} + \text{max}\{0, 2 \hat{\eta}_{t_{j}}^2\}$ and
$(\log y_{t_{j+2}} - \log y_{t_{j+1}})^2$ are independent in the additive microstructure noise
model $\log Y_{t_j} = X_{t_j} + U_{t_j}$ with $U_{t_j}$ {i.i.d.} - thus by using $(\log y_{t_{j+2}}
- \log y_{t_{j+1}})^2\;$ (\ref{CrossValidationBench}) becomes a decent estimate of the mean squared
error (plus a term constant in $\lambda$)\big). For $\hat{\eta}_{t_j}^2$ we use the step sizes
$\frac {1} {j-2}$ because $\eta_t^{2}$ should be close to a constant function.

All estimators use the true volatility as starting value. Typical outcomes of the estimators are
given in Figure~\ref{fig:SimulatedData_allEstimators}. Note that  volatility is plotted (instead of
squared volatility). In the second case (lower plot) a constant step size is clearly suboptimal.
Therefore we also computed our estimator combined with the SAGES method for adaptive step size
selection $\hat{\Sigma}_{t_{j}}^{S}$ as described in Section~\ref{ch:msn:impl}.
$\hat{\Sigma}_{t_{j}}^{S}$ is calculated using $L=15$ step sizes ranging from 0.05 to 0.00005
(equally spaced). In the first case (upper plot) the estimator $\hat{\Sigma}_{t_{j}}^{S}$ didn't
give better results than the estimator $\hat{\Sigma}_{t_{j}}$ and is therefore omitted. We also
tried to use the SAGES method for the benchmark estimator. This gave surprisingly bad results which
are not reported here.

Because the true $\Sigma(t_j)$ is known we can compute the mean squared error $\Sigma_{j=2}^{T-1}
\big(\hat{\Sigma}(t_j) - \Sigma(t_j)\big)^2$ for the estimators which gives $1.21 \times 10^{-18}$
and $1.34 \times 10^{-18}$ for $\hat{\Sigma}_{t_{j}}$ and $\hat{\Sigma}_{t_{j}}^{\text{B}}$,
respectively, for the upper plot in Figure~\ref{fig:SimulatedData_allEstimators}. For the
estimators $\hat{\Sigma}_{t_{j}}$, $\hat{\Sigma}_{t_{j}}^{S}$, and
$\hat{\Sigma}_{t_{j}}^{\text{B}}$ in the lower plot we obtain $8.59 \times 10^{-19}$, $7.52 \times
10^{-19}$, and $2.55 \times 10^{-18}$. In both plots, our estimators significantly outperforms the
benchmark estimator.

The general impression from Figure~\ref{fig:SimulatedData_allEstimators} is that the estimates are
a bit undersmoothed. As for nonparametric path-wise estimation of local volatility the noise  is to be expected if the method has a low bias, and additional smoothing will reduce variance at the expense of increased bias. We mention that additional variability comes in from the particle filter where the estimated covariance matrix is used instead of the true one.

\subsubsection*{Influence of jumps}

In particular for coarser sampling intervals there is strong evidence that stock price levels exhibit jumps - e.g. so-called rare compound Poisson jumps. Todorov and Tauchen (2011) analyze the high-frequency movements in stock market volatility using data of the VIX volatility index sampled at a 5 minute rate and even conclude that volatility should be modeled by a pure jump
process with jumps of infinite variation. Foster and Nelson (1996) acknowledge the problems that jumps may cause for local volatility estimation.

The model we have given in this paper is a model at a finer time-scale based on volatility in transaction time and trading intensity. The volatility at a larger interval of say $\triangle t = 5$ minutes would be given by $\int_{t}^{t+\Delta t}\! \Sigma (s) \, \lambda_{I}(s) \, ds$ and it is an interesting question whether part of the jumps on a larger scale can solely be explained by an increase of the trading intensity on that interval. Nevertheless jumps may also occur in $\Sigma (\cdot)$ and $\lambda_{I}(\cdot)$ - although their occurrence seems to be less frequent. A formal study on the structure of the jumps must be deferred to future work.

\begin{figure}[t]
\begin{minipage}[t]{0.60\textwidth}
\includegraphics[height=0.5\textwidth,width=\textwidth]{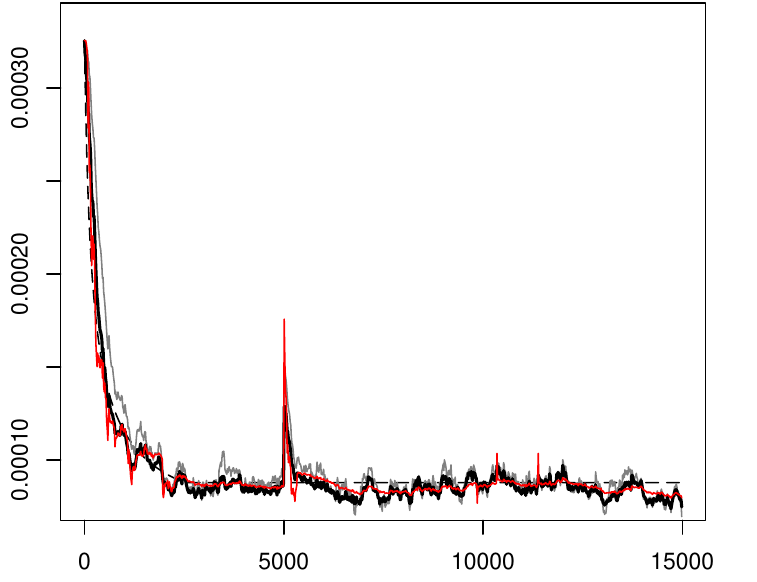}
\caption{\footnotesize Estimation of the second volatility curve from Figure~\ref{fig:SimulatedData_allEstimators} in case of a price-jump. The plot shows the
estimator $\hat{\Sigma}_{t_j}$ (black line), the adaptive version with SAGES
$\hat{\Sigma}_{t_j}^{S}$ (red (dark gray) line), and the
benchmark estimator (light gray line).}
\label{fig:SimulationJumps1}
\end{minipage} \begin{minipage}[t]{0.02\textwidth}$\quad$\end{minipage} \begin{minipage}[t]{0.37\textwidth}
\includegraphics[width=\textwidth,keepaspectratio]{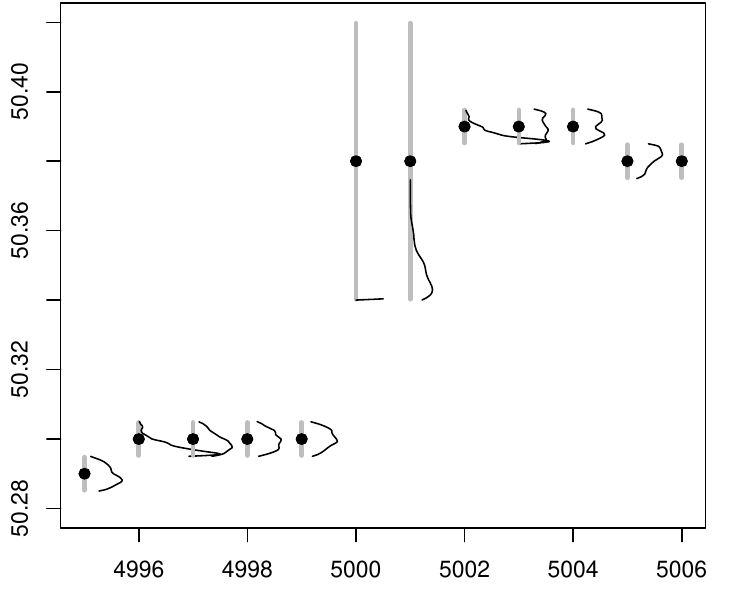}
%\vspace*{0.5cm}
\caption{\footnotesize Estimated filtering distributions about the price-jump for the market
microstructure noise model with deterministic rounding.}
\label{fig:SimulationJumps2}
\end{minipage}
\end{figure}

To investigate the influence of jumps we have taken in Figure~\ref{fig:SimulationJumps1} and \ref{fig:SimulationJumps2} the simulated data from the second plot of Figure~\ref{fig:SimulatedData_allEstimators} and added a jump of 8 cents at time 5,000 (i.e. the returns show one ``outlier'' at time 5,000).
The plot shows that the volatility estimate and the particle filter quickly recover after the jump. Furthermore we can see that the SAGES estimator recovers a bit better which is due to the adaptive stepsize selection.

\subsection{Results for Real Data} \label{ch:msn:sim:realdata}

\subsubsection*{The data and data specific modifications}

To demonstrate the method we have used stock data from the TAQ data base. Transactions and market maker quotes of the symbol C
(Citigroup) for the 3rd September 2007 were extracted from this data base. Prior to our analysis we have carried out the following obvious data cleaning steps which could also be done on-line.

\begin{description}
\item \underline{Cleaning A:} Delete all transactions (quotes) with time stamps outside the main trading
    period (9:30 AM to 4 PM).
\item \underline{Cleaning B:} Delete all transactions (quotes) that are not originating from the NYSE.
\item \underline{Cleaning C:} Delete all transactions with abnormal sale condition or corrected prices
    (see the TAQ User's Guide for details).
\end{description}
Since the time stamp precision of these data is limited to one second, several time stamps occur with multiple transactions. Since each of these transactions constitute a single step in the (transaction time) state equation (\ref{distributional_state-equation}) one should normally use all these transactions separately (e.g. with an equidistant ex post splitting of the trading times). However, a closer inspection of these multiple transactions revealed that the (time) ordering of these transactions was not preserved and we therefore decided to treat this problem like a missing data problem. This means at a time stamp with $M$ transactions and equal trading times $t_{j-M+1}= \cdots =t_{j}$ we used for the transaction time estimator $\hat{\Sigma}_{t_j}$ instead of the recursion (\ref{seqEM1h}) with $\breve{\Sigma}_{t_{j}}(\omega_{t_j})$ the corresponding $M$-step recursion with
\begin{equation*} \label{EstimateSigma4x}
\breve{\Sigma}^{M}_{t_{j}}(\omega_{t_j}) := \frac {1} {M}\sum_{i=1}^N \omega_{t_{j}}^i
\big(\mathbf{x}_{t_{j}}^i-\mathbf{x}_{t_{j-M}}^i\big)
\big(\mathbf{x}_{t_{j}}^i-\mathbf{x}_{t_{j-M}}^i\big)'
\end{equation*}
and for the classical clock time estimator $\hat{\Sigma}_{t_j}^{c}$ the recursion with
\begin{equation*} \label{}
\breve{\Sigma}_{t_j}^{cM}(\omega_{t_j}^c) := \sum_{i=1}^N \omega_{t_j}^{ci}
\frac{ \big(\mathbf{x}_{t_j}^{ci} - \mathbf{x}_{t_{j-M}}^{ci}\big) \big(\mathbf{x}_{t_j}^{ci} -
\mathbf{x}_{t_{j-M}}^{ci}\big)'} {|t_{j}-t_{j-M}|}.
\end{equation*}
For the durations the situation is different since with the number of trades the information about the trading intensity $\lambda_{I}(t)$ is (almost) fully available. We therefore apply for $k=j-M+1,\ldots,j\;$ the update
\begin{equation*} \label{}
\bar{\delta}_{k} = \big(1-  \lambda\big) \, \bar{\delta}_{k-1} +
 \lambda \, \frac {t_{j} - t_{j-M}} {M}
\end{equation*}
(i.e. $M$-times the same update - alternatively we may also use one update with $\lambda$ replaced by $M \lambda$ which for small $M \lambda$ is almost the same).

\subsubsection*{Estimation of the filtering distributions with real market maker quotes}

In order to show how our method works in the case when market maker quotes are available (Example 3 in Section~\ref{ch:msn:msnmodel}) we matched by hand (through an adjustment of the time stamps)  the quotes and transactions of symbol C for a fraction of the trading day. The particle filter is used with $N=5,000$ particles and $A_{t_j}$ as in (\ref{AtEx3}) where $\Delta_{t_j} := 0.5\, (\beta_{t_j} - \alpha_{t_j})$ to estimate the filtering distributions of the unknown efficient (log-)\,prices.
Figure~\ref{fig:MSNFilteringDistr_realQuotes_univariate} gives kernel density estimates
based on these particle
approximations. The market maker quotes, the transaction prices, and supports of the filtering
distributions are also shown. From the figure it can be seen that some filtering distributions are
highly skewed. In addition, consecutive zero returns lead to very uninformative filtering
distributions (see transactions 2,300 through 2,309).

\subsubsection*{Transaction time volatility estimation}

We apply the estimators $\hat{\Sigma}_{t_{j}}$ and $\hat{\Sigma}_{t_{j}}^{S}$ with $N=$ 500
particles and the benchmark method $\hat{\Sigma}_{t_{j}}^{\text{B}}$ from (\ref{msn:recursiveBenchTV})
to estimate the spot volatility for C. An initial volatility of $0.0005$ is used. Here we have estimated the market maker quotes from the trades, that is we have used  $A_{t_j}$ as given in (\ref{AtEx4}) first with deterministic rounding and then with stochastic rounding - see (\ref{AtEx4a}).

\begin{figure}[t]
\centering
\includegraphics[width=0.78\textwidth,keepaspectratio]{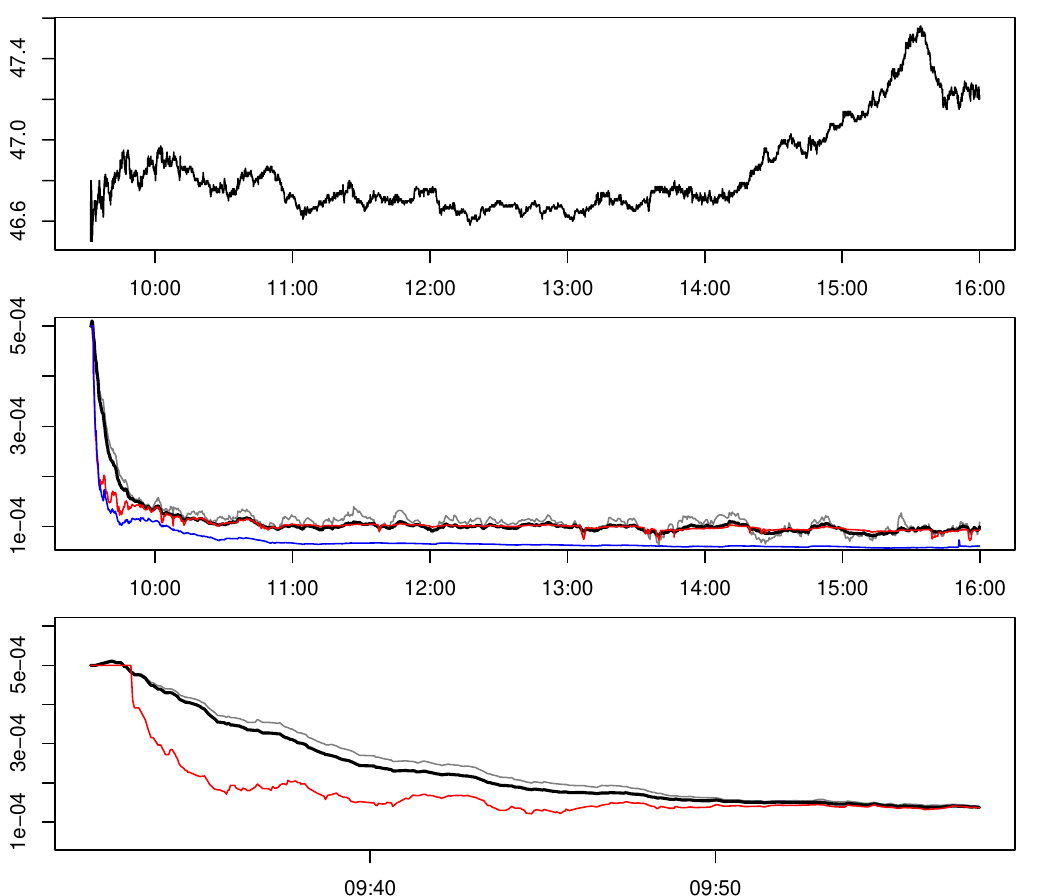}
\caption{\footnotesize
Real data example: Estimation of time-varying spot volatility in transaction
time. The upper plot shows the transaction data of the symbol
C for the 3rd September 2007. The middle and lower plot give the volatility
estimator with deterministic rounding $\hat{\Sigma}_{t_{j}}$ (black line), the adaptive version with SAGES $\hat{\Sigma}_{t_j}^{S}$ (red (dark gray) line), and the benchmark estimator $\hat{\Sigma}_{t_{j}}^{\text{B}}$ (light gray
line). The blue (lower gray) line in the middle plot shows the estimator $\hat{\Sigma}_{t_{j}}$ with stochastic rounding.}
\label{fig:RealData_allEstimatorsStartValue}
\end{figure}

The transaction data of C and the volatility estimators are shown in
Figure~\ref{fig:RealData_allEstimatorsStartValue}. At the beginning of the trading day the
volatility is large and highly varying. Later, the volatility settles down and seems to be almost
constant. Therefore, the SAGES method for localized step size selection is advantageous
compared to fixed step sizes. Again the benchmark estimator is rougher than our estimators.
Practically, the \underline{transaction time volatility} is almost constant after
11:00~am which in our experience is a typical feature of transaction data of liquid stocks. Contrary to this the \underline{clock time volatility} is more fluctuating and shows well known features like the U-shape. This has already been discussed at the end of Section~\ref{ch:msn:clock}.

The blue (lower gray) line in Figure~\ref{fig:RealData_allEstimatorsStartValue} shows the estimator $\hat{\Sigma}_{t_{j}}$ with stochastic rounding. The difference to the black estimator with deterministic rounding is quite large - but can be explained heuristically: In some sense all methods decompose the realized volatility into the ``true'' volatility and the volatility coming from  microstructure noise. Since the microstructure noise model with stochastic rounding has higher volatility than the deterministic one it is obvious that the resulting volatility of the unobserved efficient price must be smaller. It is remarkable that the first estimator has the same level as the benchmark-estimator (which uses a completely different linear microstructure noise model). In our opinion this is another indicator that the microstructure noise model with deterministic rounding is preferable to the model with stochastic rounding.

\subsubsection*{Clock time spot volatility estimation}

The corresponding clock time volatility estimators have already been displayed in Figure~\ref{fig:RealDataClock_allEstimators} where also the transition from transaction time to clock time has been discussed. In Figure~\ref{fig:RealDataClock_allEstimators} all estimators have been calculated with the SAGES-method. Since volatility in clock time is more volatile than in transaction time here SAGES requires larger step sizes. We use step sizes equally spaced between 0.3 and 0.003.

\begin{figure}[t]
\centering
\includegraphics[width=0.78\textwidth,keepaspectratio]{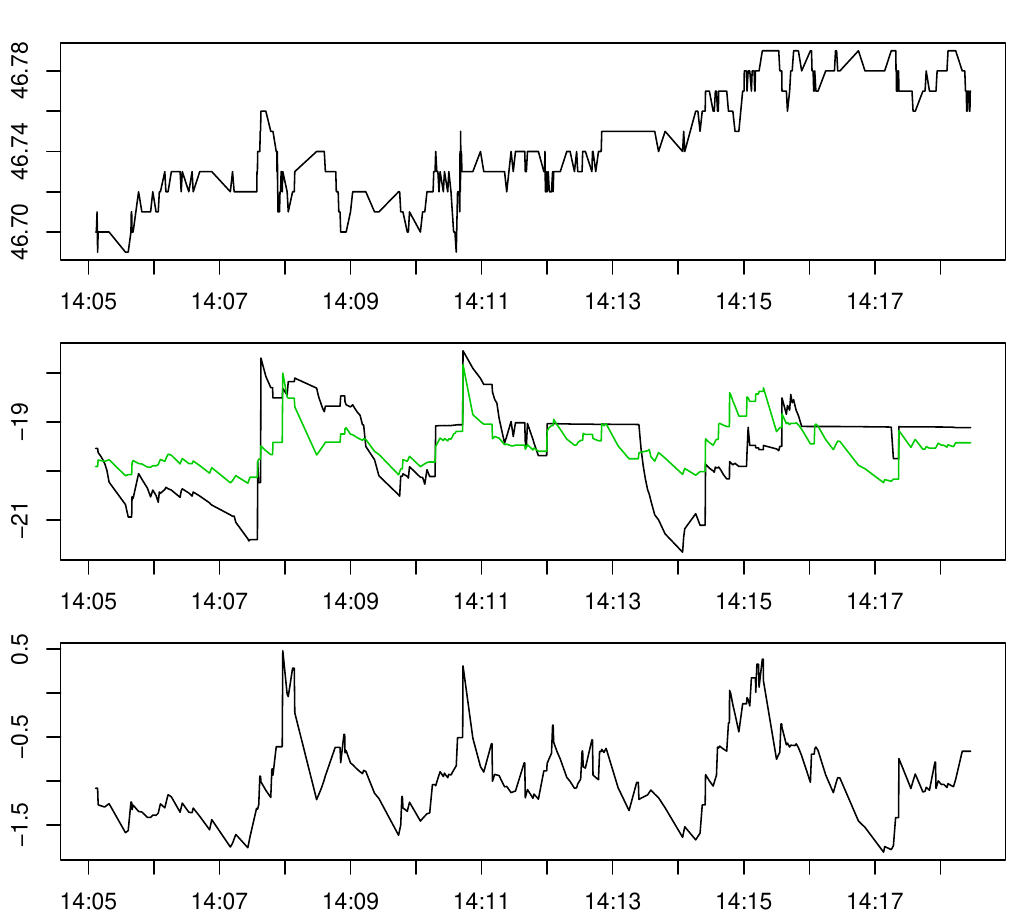}
\caption{\footnotesize
Results from Figure~\ref{fig:RealDataClock_allEstimators} for a small fraction of the trading day. The upper plots show the transaction prices of C; the middle plot gives the log of the volatility estimators $\hat{\Sigma}^{cS}_{t_j}$ (black line) and $\hat{\Sigma}_{\text{alt}}^{cS}(t_j)$ (green (gray) line); the lower plot shows the log of the trading intensity $\hat{\lambda}_{I}(t_j)$.}
\label{fig:RealDataClock_allEstimatorsA}
\end{figure}

For the duration estimator $\bar{\delta}_{j}$ we determined the stepsize by minimizing the prediction error $\Sigma_{j=2}^{T-1} \big\{\bar{\delta}_{j} - (t_{j+1}-t_{j})\big\}^2$ leading to approximately $\lambda=0.08$. (Because of the dependence of the durations, $\bar{\delta}_{j}$ and $(t_{j+1}-t_{j})$ usually are not independent and minimization of the above criterion therefore is not approximately the same as minimization of the mean squared error. Despite of this we think that the resulting $\lambda$ is reasonable.)

In addition to Figure~\ref{fig:RealDataClock_allEstimators} above Figure~\ref{fig:RealDataClock_allEstimatorsA} compares the transaction data and the volatility estimates for a small time period. Note that the estimator needs about one minute to settle down again after the
occurrence of a spike.

\section{Concluding Remarks} \label{ch:msn:concl}

We have presented a technique for the on-line estimation of time-varying volatility based on
noisy transaction data. The algorithm updates the volatility estimate immediately after a new transaction. On a recent personal computer an efficient implementation of the method requires a few milliseconds for a single update of the estimator (including one iteration of the particle filter with 500 particles).

The paper contains different contributions: First, we have proposed a nonlinear market
microstructure noise model that covers bid-ask bounces, time-varying bid-ask spreads, and the
discreteness of prices observed in real data. Second, the problem of on-line volatility estimation
has been treated in a nonlinear state-space framework. The filtering distribution of the efficient
price is approximated with a particle filter and the volatility is estimated as a parameter
of the filtering distribution. Third, we have presented a sequential EM-type algorithm which
allows the on-line estimation of time-varying volatility.

We also make a clear distinction between the (spot)
volatility per time unit $\Sigma^{c}(t)$ and the volatility per transaction $\Sigma(t)$. We have  used a diffusion model with random time change given by the total number of transactions. This leads to a decomposition of volatility in clock time into volatility in transaction time and trading intensity. At least for our data set it turned out that volatility in transaction time is almost constant (after some steep decrease at the beginning of the trading day), and the fluctuation of clock time volatility is merely a result of the fluctuation of the trading intensity. In our data also the increase of volatility in the afternoon (part of the U-shape) is a feature of the trading intensity and not of the transaction time volatility.

We mention that most components of this method can be used in combination with other models or estimation techniques: For example the particle filter can be used with other price models (e.g. with a drift term) or other microstructure noise models. Likewise the decomposition of clock time volatility into transaction time volatility and trading intensity can be used with linear microstructure noise models and other estimation techniques.

Of course it is desirable to have a complete mathematical theory for the methods of this paper. However, we think that this is very hard to achieve. Mathematically exact are the results on the particle filter given that the true volatility is known (i.e. with $\hat{\Sigma}_{t_{j}}^{\text{pf}} = \Sigma_{t_{j}}$) - in particular
the results from Proposition~\ref{msn:prop1} on the optimal proposal and the importance weights.
This means that the particle filter determines correctly the conditional distribution of the
efficient prices given the observations. In the simpler
context of {i.i.d.}-observations convergence properties of recursive EM-type algorithms have been
studied in Titterington (1984), Sato (2000), Wang and Zhao (2006). Capp{\'{e}} and Moulines (2009)
derive asymptotic normality with rate of convergence $\lambda^{1/2}$ for a similar recursive EM-type algorithm in an i.i.d setting. In the present situation we may hope for a similar result provided that the model in (\ref{roundssm:synchObsEquation}) and (\ref{roundssm:synchStateEquation}) is properly rescaled with volatility $\Sigma(t/T)$ and the curve $\Sigma(\cdot)$ is sufficiently smooth. A similar result can be found in Dahlhaus and Subba Rao (2007) where the asymptotic properties of a recursive ARCH-estimator have been derived. The optimal rate of convergence will however not be attained since the recursive estimator is one-sided. The data-adaptive SAGES-procedure will make it even more difficult to derive the asymptotic distribution. For that reason we recommend a simulation based on the estimated volatility curve for deriving approximate confidence intervals.

\medskip

\noindent  \textbf{Acknowledgement:} We are very grateful to the Co-Editor Professor George Tauchen  and an anonymous referee whose comments helped to improve the paper considerably.

\medskip

\noindent  \textbf{Disclaimer:} The views expressed here are those of the authors and not
necessarily those of its employers.

\section*{References}

\begin{description}
\baselineskip1.3em
\itemsep-0.04cm
\item A\"{i}t-Sahalia, Y., Mykland, P.A.,  and Zhang, L. (2005) How Often to Sample a
    Continuous-Time Process in the Presence of Market Microstructure Noise. {\it Review of
    Financial Studies}, 18, 351-416.

\item Andersen, T.G., Bollerslev, T., and Meddahi, N. (2006) Realized Volatility Forecasting
    and Market Microstructure Noise. unpublished manuscript.

\item Andrieu, C., and Doucet, A. (2002) Particle Filtering for partially observed Gaussian
state space models. {\it Journal of the Royal Statistical Society B}, 64, 827-836.

\item An{\'{e}}, T., and Geman, H. (2000) Order Flow, Transaction Clock, and Normality of Asset
    Returns. {\it The Journal of Finance}, 55, 2259-2284.

\item Ball, C.A. (1988) Estimation Bias Induced by Discrete Security Prices. {\it The Journal
    of Finance}, 43, 841-865.

\item Bandi, F.M., and Russell, J.R. (2006) Seperating microstructure noise from volatility.
    {\it Journal of Financial Economics}, 79, 655-692.

\item --- (2008) Microstructure noise, realized variance, and optimal sampling. {\it Review of
    Economic Studies}, 75, 339-369.

\item Barndorff-Nielsen, O.E., Hansen, P.R., Lunde, A., and Shephard, N. (2008) Designing
    Realized Kernels to Measure the Ex-Post Variation of Equity Prices in the Presence of
    Noise. {\it Econometrica}, 76, 1481-1536.

\item Bos, C.S., Janus, P., and Koopman, S.J. (2009) Spot Variance Path Estimation and its
    Application to High Frequency Jump Testing. Discussion Paper TI 2009-110/4, Tinbergen
    Institute.

\item Capp{\'{e}}, O., and Moulines, E. (2009) On-line expectation-maximization algorithm for
    latent data models. {\it Journal of the Royal Statistical Society, Series B}, 71, 593-613.

\item Chen, Y. and Spokoiny, V. (2009) Modeling and estimation for nonstationary time series with
applications to robust risk management. unpublished manuscript.

\item Christensen, K., Podolskij, M., and Vetter, M. (2009) Bias-correcting the realised
    range-based variance in the presence of market microstructure noise. {\it Finance and
    Stochastics}, 13, 239-268.

\item Clark, P. (1973) A subordinated stochastic process model with finite variance for speculative prices. {\it Econometrica}, 41, 135-155.

\item Dahlhaus R. and Subba Rao S. (2007) A recursive online algorithm for the estimation of time-varying ARCH parameters. {\it Bernoulli}, 13, 389-422.

\item Delattre, S. and Jacod, J. (1997). A central limit theorem for normalized functions of the increments of a diffusion process, in the presence of round-off errors. {\it Bernoulli}, 3, 1�28.

\item Dempster, A.P., Laird, N.M., and Rubin, D.B. (1977) Maximum Likelihood from Incomplete
    Data via the EM Algorithm. {\it Journal of the Royal Statistical Society, Series B}, 39,
    1-38.

\item Douc, R., Capp{\'{e}}, O., and Moulines, E. (2005) Comparison of resampling schemes for
    particle filtering. In {\it Proceedings of the 4th International Symposium on Image and
    Signal Processing and Analysis}, 64-69.

\item Doucet, A., Godsill, S., and Andrieu, C. (2000) On sequential Monte Carlo sampling
    methods for Bayesian filtering. {\it Statistics and Computing}, 10, 197-208.

\item Doucet, A., de Freitas, N., and Gordon, N. (ed.) (2001) {\it Sequential Monte Carlo
    Methods in Practice}. New York: Springer.

\item Even-Dar, E., and Mansour, Y. (2003) Learning Rates for Q-learning. {\it Journal of
    Machine Learning Research}, 5, 1-25.

\item Fan, J., and Wang, Y. (2008) Spot volatility estimation for high-frequency data. {\it
    Statistics and Its Interface}, 1, 279-288.

\item Foster, D., and Nelson, D. (1996) Continuous Record Asymptotics for Rolling Sample
    Estimators. {\it Econometrica}, 64, 139-174.

\item Gabaix, X., Gopikrishnan, P., Plerou, V., and Stanley, H.E. (2003) A theory of power-law
    distributions in financial market fluctuations. {\it Nature}, 423, 267-270.

\item Genz, A. (1992) Numerical computation of multivariate normal probabilities. {\it Journal
    of Computational and Graphical Statistics}, 1, 141-149.

\item Genz, A. (2004) Numerical computation of rectangular bivariate and trivariate normal and
    t probabilities. {\it Statistics and Computing}, 14, 151-160.

\item Geweke, J. (1991) Efficient simulation from the multivariate normal and student-t
    distributions subject to linear constraints. {\it In Computing Science and Statistics:
    Proceedings of the 23rd Symposium on the Interface, Ed. E. Keramidas and S. Kaufman},
    571-578. American Statistical Association, Alexandria, VA.

\item Hansen, P.R., and Lunde, A. (2006) Realized Variance and Market Microstructure Noise.
    {\it Journal of Business and Economics Statistics}, 24, 127-161.

\item Hansen, P.R. and Horel, G. (2009). Quadratic Variation by Markov Chains. CREATES Research Paper 2009-13.

\item Harris, L. (1990) Estimation of Stock Price Variances and Serial Covariances from
    Discrete Observations. {\it Journal of Financial and Quantitative Analysis}, 25, 291-306.

\item Hasbrouck, J. (1999) Security Bid/Ask Dynamics with Discreteness and Clustering. {\it Journal of Financial Markets}, 2, 1-28.

\item Hasbrouck, J. (2004) Liquidity in the Futures Pits: Inferring Market Dynamics
		from Incomplete Data. {\it Journal of Financial and Quantitative Analysis}, 39, 2.

\item Howison, S., and Lamper, D. (2001) Trading volume in models of financial derivatives.
    {\it Applied Mathematical Finance}, 8, 119-135.

\item Jacod, J., Li, Y., Mykland, P.A., Podolskij, M., and Vetter, M. (2009) Microstructure
    noise in the continuous case: The pre-averaging approach. {\it Stochastic Processes and
    their Applications}, 119, 2249-2276.

\item Joe, H. (1995) Approximations to multivariate normal rectangle probabilities based on
    conditional expectations. {\it Journal of the American Statistical Association}, 90,
    957-964.

\item Kalnina, I., and Linton, O. (2008) Estimating quadratic variation consistently in the
    presence of endogenous and diurnal measurement error. {\it Journal of Econometrics}, 147,
    47-59.

\item Kong, A., Liu, J., and Wong, W. (1994) Sequential imputation and Bayesian missing data
    problems. {\it Journal of American Statistical Association}, 89, 278-288.

\item Kotecha, J. and Djuric, P. (1999) Gibbs sampling approach for the
		generation of truncated multivariate Gaussian random variables. {\it Proceedings of the IEEE International
    Conference on Acoustics, Speech and Signal Processing}, 1757-1760.

\item Kristensen, D. (2010) Nonparametric filtering of the realized spot volatility: A
    Kernel-based Approach. {\it Econometric Theory}, 26, 60-93.

\item Large, J. (2007) Estimating Quadratic Variation When Quoted Prices Change By A Constant
    Increment. unpublished manuscript.

\item Li, Y., and Mykland, P.A. (2007) Are volatility estimators robust with respect to
    modeling assumptions?. {\it Bernoulli}, 13, 601-622.

\item Manrique, A., and Shephard, N. (1998) Simulation-based likelihood inference for
		limited dependent processes. {\it Econometrics Journal}, 1, C174-C202.

\item Munk, A., and Schmidt-Hieber, J. (2009) Nonparametric Estimation of the Volatility
    Function in a High-Frequency Model corrupted by Noise. unpublished manuscript.

\item  Owens, J.P. and Steigerwald, D.G. (2006) Noise reduced realized volatility: a
Kalman filter approach. {\it Advances in Econometrics} 20 (ed. Tom Fomby and Dek
Terrell), 211-227, Elsevier.

\item Plerou, V., Gopikrishnan, P., Gabaix, X., A Nunes Amaral, L., and Stanley, H.E. (2001)
    Price fluctuations, market activity and trading volume. {\it Quantitative Finance}, 1,
    262-269.

\item Podolskij, M., and Vetter, M. (2009) Estimation of volatility functionals in the
    simultaneous presence of microstructure noise and jumps. {\it Bernoulli}, 15, 634-658.

\item Robert, C. (1995) Simulation of truncated normal variables. {\it Statistics and
    Computing}, 5, 121-125.

\item Robert, C.Y., and Rosenbaum, M. (2008) Ultra high frequency volatility and co-volatility
    estimation in a microstructure model with uncertainty zones. unpublished manuscript.

\item Rodriguez-Yam, G., Davis, R., and Scharf, L. (2004) Efficient gibbs sampling of truncated
    multivariate normal with application to constrained linear regression. Technical report,
    Colorado State University, 2004.

\item Rosenbaum, M. (2009) Integrated volatility and round-off error. {\it Bernoulli}, 15,
    687-720.

\item Sato, M. (2000) Convergence of on-line EM algorithm. In {\it Proc. Int. Conf. on Neural
    Information Processing}, 1, 476-481.

\item Todorov, V. and Tauchen, G. (2011) Volatility Jumps. {\it J. Business and Economic Statistics}, 29, 356-371.

\item Titterington, D.M. (1984) Recursive Parameter Estimation Using Incomplete Data. {\it
    Journal of the Royal Statistical Society, Series B}, 46, 257-267.

\item Voev, V., and Lunde, A. (2007) Integrated Covariance Estimation using High-Frequency Data
    in the Presence of Noise. {\it Journal of Financial Econometrics}, 5, 68-104.

\item Wang, S., and Zhao, Y. (2006) Almost sure convergence of Titterington's recursive
    estimator for mixture models. {\it Statistics \& Probability Letters}, 76, 2001-2006.

\item Zeng, Y. (2003) A Partially Observed Model for Micromovement of Asset Prices with Bayes
    Estimation via Filtering. {\it Mathematical Finance}, 13, 411-444.

\item Zhang, L., Mykland, P.A., and A\"{i}t-Sahalia (2005) A Tale of Two Time Scales:
    Determining Integrated Volatility with Noisy High-Frequency Data. {\it Journal of the
    American Statistical Association}, 100, 1394-1411.

\item Zhou, B. (1996) High-Frequency Data and Volatility in Foreign-Exchange Rates. {\it
    Journal of Business \& Economic Statistics}, 14, 45-52.

\item Zu, Y. and Boswijk, P. (2010) Estimating spot volatility with high frequency
     financial data. Preprint, University of Amsterdam.

\item Zumbach, G, Corsi, F., and Trapletti, A. (2002) Efficient estimation of volatility using
    high-frequency data. Technical Report, Olsen \& Associates.

\end{description}

\end{document}